\documentclass[10pt,aps,prb,twocolumn,eqsecnum]{revtex4}
\usepackage[active]{srcltx}
\usepackage[utf8]{inputenc}
\usepackage{latexsym}
\usepackage{amsmath,amssymb,bbm,braket,bm}
\usepackage{graphicx}
\usepackage[hypertex]{hyperref}
\usepackage{ulem}
\usepackage{color}

\def\:={\,\raisebox{0.85pt}{.}\hspace{-2.78pt}\raisebox{2.85pt}{.}\!\!=\,}
\def\=:{\,=\!\!\raisebox{0.85pt}{.}\hspace{-2.78pt}\raisebox{2.85pt}{.}\,}

\begin{document}

\title{Effective field theory for the bulk-edge correspondence
in a two-dimensional $\mathbb{Z}^{\,}_{2}$ topological insulator
with Rashba interactions}

\author{Pedro R. S. Gomes}
\affiliation{Department of Physics, Boston University\\
Boston, MA, 02215, USA}

\author{Po-Hao Huang}
\affiliation{Department of Physics, Boston University\\
Boston, MA, 02215, USA}

\author{Claudio Chamon}
\affiliation{Department of Physics, Boston University\\
Boston, MA, 02215, USA}

\author{Christopher Mudry}
\affiliation{Condensed Matter Theory Group, Paul Scherrer Institute\\
CH-5232 Villigen PSI, Switzerland}

\begin{abstract}

We determine the effective field theory in $(2+1)$-dimensional space
and time that it captures the long-wave-length and low-energy limit of
fermions hopping on a honeycomb lattice at half-filling
when both a dominant intrinsic and subdominant
Rashba spin-orbit couplings are present. This effective field theory for a
$\mathbb{Z}^{\,}_{2}$ topological insulator
(the Kane-Mele model at vanishing uniform and staggered chemical potentials)
is a perturbation around a double Chern-Simons theory,
with the $U(1)$ gauge invariance associated to spin conservation
explicitly broken due to the Rashba spin orbit coupling. Nonetheless,
we find that the effective field theory has a BRST symmetry
that allows us to construct the bulk-edge correspondence.

\end{abstract}
\maketitle

\section{Introduction}
\label{sec: Introduction}

There has been a great amount of interest in the
field of research
opened up by the discovery of materials known as
topological insulators.~\cite{Hasan,Qi_Zhang,Ando} Topological insulators
represent a new quantum state of matter that is characterized by
bulk properties like those of ordinary band
insulators, but supporting protected conducting boundary states
on their edges or surfaces. These states are possible due to a
combination of spin-orbit interactions and time-reversal symmetry.

The Kane-Mele model introduced in Refs. ~\onlinecite{Kane1} and
\onlinecite{Kane2} is an example of a band insulator in
two-dimensional space for which time-reversal symmetry guarantees the
stability of gapless edge states that are perfectly conducting along
any boundary.  The Kane-Mele model is a tight-binding representation
for electrons in graphene in the presence of an intrinsic spin-orbit
coupling and of a Rashba spin-orbit coupling.  Even though the
magnitudes of the spin-orbit couplings in graphene are too small to
lead to observable effects with the present experimental resolution in
energy and temperature, the Kane-Mele model aroused considerable
interest and led to the predictions and discoveries of
$\mathbb{Z}^{\,}_{2}$ topological insulators both in two- and
three-dimensional space (see Refs.~\onlinecite{Hasan},
\onlinecite{Qi_Zhang}, and~\onlinecite{Ando}
for reviews).

The Kane-Mele model at vanishing uniform and staggered chemical potentials and
in the absence of the Rashba spin-orbit coupling
simplifies to a reducible massive Dirac Hamiltonian with
Dirac matrices of rank 8 at long wave lengths and low energies.
In turn, each irreducible block realizes a massive Dirac Hamiltonian with
Dirac matrices of rank 4. There are thus two
Dirac masses that enter with opposite signs
so that time-reversal symmetry holds.
Both the electronic charge and the projection of the electronic
spin quantum number along the quantization axis in spin space are conserved
when the Rashba terms are switched off in the Kane-Mele Hamiltonian.
Integration of the electrons in the Kane-Mele Hamiltonian
at vanishing uniform and staggered chemical potentials,
without Rashba terms, but coupled to two $U(1)$ external gauge fields,
one that couples to the conserved $U(1)$ charge and one that couples
to the $U(1)$ spin current, delivers a double Chern-Simons (CS) theory.\
\cite{Vozmediano}
As there is a bulk-edge correspondence associated to each of the CS
terms, there follows the existence and stability of an integer number of pairs of helical edge states in any geometry with boundaries.
Correspondingly, the Kane-Mele model in the absence of the Rashba terms
supports the quantum-spin Hall effect.\cite{Bernevig}
The quantization of the spin Hall response is lost for any Rashba
spin-orbit coupling. The insight of Kane and Mele was to recognize
that, as long as time-reversal symmetry holds, a single pair of helical edge states
persists in the form of a perfectly
conducting channel, provided there was an odd number of pairs
of helical edge states prior to switching on the (not too large)
Rashba spin-orbit coupling.

The goal of this work is to derive the effective quantum field theory
in $(2+1)$-dimensional space and time that encodes at long wave
lengths and low energies the Kane-Mele model
at vanishing uniform and staggered chemical potentials together
with an intrinsic spin-orbit coupling that dominates over a Rashba spin-orbit
coupling, and understand how the gapless edge dynamics arises from
this bulk action. A brief summary with the main results
of the paper follows.

Starting with a Dirac Hamiltonian coupled to the pair
$A^{(+)}_{\mu}$
and
$A^{(-)}_{\mu}$
of gauge fields and after integrating out the massive
Dirac fermions, we obtain the one-loop effective action
\begin{subequations}
\begin{widetext}
\begin{equation}
I[A^{(+)},A^{(-)}]=
-\frac{1}{4\pi}
\int\mathrm{d}^{3}x
\left[
\epsilon^{\mu\alpha\nu}\,
\left(
A^{(+)}_{\mu}\partial^{\,}_{\alpha}A^{(-)}_{\nu}
+
A^{(-)}_{\mu}\partial^{\,}_{\alpha}A^{(+)}_{\nu}
\right)
-
\frac{\lambda^{2}_{\mathrm{R}}}{|\eta|}
\left(A^{(-)}_{0}
\right)^{2}
\right],
\label{eq: first main result of the paper}
\end{equation}
\end{widetext}
where the real-valued parameters $\eta$ and
$\lambda^{\,}_{\mathrm{R}}$
are the spin-orbit and Rashba couplings, respectively.
This action is invariant under
gauge transformations of the field $A^{(+)}_{\mu}$
as charge is conserved. It is not invariant
under gauge transformations of the field $A^{(-)}_{\mu}$
as the spin-1/2 symmetry is completely broken by the
Rashba spin-orbit coupling.
As it is known for the quantum Hall effect,\
\cite{Frohlich1,Frohlich2,Wen}
gauge invariance is sufficient
to show the existence of the propagating chiral
states along the edge.
The question we are thus after
is how to construct the bulk-edge correspondence
without the complete $U(1)\times U(1)$ gauge invariance
of the effective action when $\lambda^{\,}_{\mathrm{R}}=0$.
The important point is that we can interpret the correction
$(A^{(-)}_{0})^{2}$ as a gauge fixing term.
By using the Faddeev-Popov procedure,~\cite{Faddeev-Popov}
we can introduce the ghost action
\begin{equation}
S^{\,}_{\mathrm{ghost}}\:=
-\frac{1}{4\pi}
\int\mathrm{d}^{3}x\,\bar{C}\,\partial^{\,}_{t}C,
\end{equation}
where $C$ and $\bar{C}$ are fermionic ghosts fields,
such that the complete action
\begin{equation}
S\:=I+S^{\,}_{\mathrm{ghost}}
\end{equation}
\end{subequations}
changes by a total derivative
under the combination of
the usual gauge transformation for the $A^{(+)}_{\mu}$ field,
\begin{subequations}
\begin{equation}
A^{(+)}_{\mu}\rightarrow
A^{(+)}_{\mu}+\partial^{\,}_{\mu}\Lambda^{(+)},
\end{equation}
with the BRST transformations~\cite{BRS,T,Weinberg,Polchinski} for
$A^{(-)}_{\mu}$, $C$, and $\bar{C}$,
\begin{equation}
\begin{split}
&
A^{(-)}_{\mu}\rightarrow
A^{(-)}_{\mu}
+
\theta\,
\partial^{\,}_{\mu}C,
\\
&
C\rightarrow
C,
\\
&
\bar{C}\rightarrow
\bar{C}
+
2\,
\frac{\lambda^{2}_{\mathrm{R}}}{|\eta|}\,
\theta\,
A^{(-)}_{0},
\end{split}
\end{equation}
\end{subequations}
where $\theta$ is a constant Grassmann-valued parameter.  Notice that
when $\lambda^{\,}_{\mathrm{R}}\rightarrow 0$, the ghosts do not
change and the transformation of the gauge field $A^{(-)}_{\mu}$
reduces to a usual gauge transformation with parameter
$\Lambda^{(-)}\equiv \theta\,C$. For any
manifold with boundaries, imposing the symmetry under
this $U(1)\times\text{BRST }$ is sufficient to derive
the bulk-edge correspondence, as will be shown
later.

The paper is organized as follows.
In Sec.\ \ref{sec: The model},
we introduce the model and formulate the problem
in a field theory form.
In Sec.\ \ref{sec: Perturbative Rashba Coupling},
we perform the one-loop calculation of the gauge effective action.
Section \ref{sec: Edge Theory}
is dedicated to the study of the edge theory.
A summary and additional comments are
presented in the Sec.\ \ref{sec: Discussions}
Three appendices contain
further details of some calculations.

\section{The model}
\label{sec: The model}

\subsection{Hamiltonian}

In this work, we consider the single-particle
Kane-Mele Hamiltonian in the Dirac approximation.
In other words, we start from the tight-binding
Hamiltonian for graphene perturbed by
an intrinsic spin-orbit coupling and a Rashba spin-orbit coupling.
At half-filling, the dispersion of graphene, to linear order in
a gradient expansion in the deviations about the Fermi momenta,
is that of an 8-dimensional representation of the
massless Dirac Hamiltonian in two-dimensional space. The intrinsic
spin-orbit coupling is represented by a mass term in the Dirac
approximation. Unlike the spin-orbit coupling, the
Rashba spin-orbit coupling is represented by an element of the
Clifford algebra that does not anticommute with the
kinetic energy. The resulting second-quantized Hamiltonian
\begin{subequations}
\begin{equation}
H\:=
H^{\,}_{0}
+
H^{\,}_{\mathrm{gauge}}
+
H^{\,}_{\mathrm{SO}}
+
H^{\,}_{\mathrm{R}}
\label{s1.1}
\end{equation}
comprises four quadratic terms in the creation and annihilation
operators obeying the fermion algebra. There is the kinetic energy
\begin{equation}
H^{\,}_{0}\:=
\psi^{\dag}
\left(
\begin{array}{cc}
\displaystyle
-\mathrm{i}\alpha^{\,}_{i}\,\partial^{\,}_{i}
&
\displaystyle
0
\\
\displaystyle
0
&
\displaystyle
-\mathrm{i}\alpha^{\,}_{i}\,\partial^{\,}_{i}
\\
\end{array}
\right)\,
\psi,
\label{s1.2}
\end{equation}
where the Latin index $i=1,2$ is reserved for the space coordinates
and the summation convention over repeated indices is
assumed. There is the coupling (the coupling $e$ is real valued)
\begin{equation}
H^{\,}_{\mathrm{gauge}}\:=
e\,
\psi^{\dag}
\left(
\begin{array}{cc}
\displaystyle
\alpha^{\,}_{i}\,
A^{u}_{i}
&
\displaystyle
0
\\
\displaystyle
0
&
\displaystyle
\alpha^{\,}_{i}\,
A^{d}_{i}
\\
\end{array}
\right)\,
\psi
\label{s1.3}
\end{equation}
to the independent pair of classical vector gauge fields
$A^{u}_{i}$ and $A^{d}_{i}$. There is the intrinsic spin-orbit coupling
(the coupling $\eta$ is real valued)
\begin{equation}
H^{\,}_{\mathrm{SO}}\:=
\mathrm{i}\eta\,
\psi^{\dag}\,
\alpha^{\,}_{1}\,
\alpha^{\,}_{2}
\otimes
s^{\,}_{3}\,
\psi
\label{s1.4}
\end{equation}
that anticommutes with
$H^{\,}_{0}$ and $H^{\,}_{\mathrm{gauge}}$. There is the Rashba
spin-orbit coupling (the coupling $\lambda^{\,}_{\mathrm{R}}$ is real valued)
\begin{equation}
H^{\,}_{\mathrm{R}}\:=
\lambda^{\,}_{\mathrm{R}}\,
\psi^{\dag}\,
\left(
\alpha^{\,}_{1}\otimes s^{\,}_{2}
-
\alpha^{\,}_{2}\otimes s^{\,}_{1}
\right)\,
\psi.
\label{s1.5}
\end{equation}
In these expressions, $\psi$ denotes the 8-component operator-valued spinor
\begin{equation}
\psi\:=
\left(
\begin{array}{cc}
\displaystyle
\psi^{u}
\\
\displaystyle
\psi^{d}
\\
\end{array}
\right),
\qquad
\psi^{u,d}\:=
\left(
\begin{array}{cc}
\displaystyle \psi^{u,d}_{+A} \\
\displaystyle \psi^{u,d}_{+B} \\
\displaystyle \psi^{u,d}_{-B} \\
\displaystyle \psi^{u,d}_{-A} \\
\end{array}
\right),
\end{equation}
where the index $u$ ($d$) refers to the spin up (down) projection
along the spin-$1/2$ quantization axis of the electrons in graphene
selected by the intrinsic spin-orbit coupling,
the indices $A$ and $B$ represent the two
sublattices of the honeycomb lattice of graphene, and the indices
$+$ and $-$ refer to the two Dirac points of graphene at half-filling.
Finally, the Dirac matrices $\alpha^{\,}_{i}$ with $i=1,2$ are chosen to be
\begin{equation}
\alpha^{\,}_{i}\:=
\sigma^{\,}_{3}\otimes\tau^{\,}_{i},
\qquad
\alpha^{\,}_{3}\:=
\sigma^{\,}_{3}\otimes\tau^{\,}_{3},
\label{s1.6}
\end{equation}
\end{subequations}
where
$\sigma^{\,}_{\mu}\equiv(\sigma^{\,}_{0},\bm{\sigma})$,
$\tau^{\,}_{\mu}\equiv(\tau^{\,}_{0},\bm{\tau})$, and
$s^{\,}_{\mu}\equiv(s^{\,}_{0},\bm{s})$
each represent three independent sets of the Pauli matrices
augmented by the unit $2\times2$ matrices.

Alternatively, we may choose to quantize the theory with
a path integral over the independent Grassmann-valued
spinors $\bar{\psi}$ and $\psi$ weighted by a Boltzmann weight
with the Lagrangian density
\begin{subequations}
\begin{align}
\mathcal{L}\:=&\,
\bar{\psi}^{u}
\left(
\mathrm{i}
\slash\!\!\!{\partial}
-
e\slash\!\!\!\!{A^{u}}
-
\eta\gamma^{\,}_{5}\,
\gamma^{3}
\right)\,
\psi^{u}
\nonumber\\
&\,
+
\bar{\psi}^{d}\,
\left(
\mathrm{i}\slash\!\!\!{\partial}
-
e\,\slash\!\!\!\!{A^{d}}
+
\eta\,
\gamma^{\,}_{5}\,
\gamma^{3}
\right)\,
\psi^{d}
\nonumber\\
&\,
+
\lambda^{\,}_{\mathrm{R}}\,
\bar{\psi}^{u}\,
\left(
-\mathrm{i}\gamma^{1}
-
\gamma^{2}
\right)\,
\psi^{d}
+
\lambda^{\,}_{\mathrm{R}}\,
\bar{\psi}^{d}\,
\left(
\mathrm{i}\gamma^{1}
-
\gamma^{2}
\right)\,
\psi^{u},
\label{s1.7}
\end{align}
where $\bar{\psi}^{u,d}\equiv (\psi^{u,d})^{\dagger}\gamma^0 $, $\slash\!\!\!\!{A}\equiv \gamma^{\mu}\,A^{\,}_{\mu}$ with the
summation convention implied over the repeated index
$\mu=0,1,2$, and the Dirac matrices $\gamma^{\mu}$ are defined by
\begin{equation}
\begin{split}
&
\gamma^{0}\equiv
\beta\:=
\sigma^{\,}_{1}\otimes\tau^{\,}_{0},
\\
&
\gamma^{1}\:=
\beta\,\alpha^{\,}_{1},
\qquad
\gamma^{2}\:=
\beta\,\alpha^{\,}_{2},
\qquad
\gamma^{3}\:=
\beta\,
\alpha^{\,}_{3},
\\
&
\gamma^{\,}_{5}\equiv
\gamma^{5}\equiv
-\mathrm{i}
\alpha^{\,}_{1}\,
\alpha^{\,}_{2}\,
\alpha^{\,}_{3}=
\mathrm{i}\gamma^{0}\,
\gamma^{1}\,
\gamma^{2}\,
\gamma^{3}=
\sigma^{\,}_{3}\otimes\tau^{\,}_{0}.
\label{s1.8}
\end{split}
\end{equation}
\end{subequations}
By using the fact that $\bar\psi$ and $\psi$ are independent
Grassmann-valued spinors, this Lagrangian density is brought
to a more convenient form by
introducing the spinors $\bar\chi^{u,d}$ and $\chi^{u,d}$ through
\begin{subequations}
\begin{equation}
\bar{\psi}^{u,d}\=:
\bar{\chi}^{u,d}\,
\gamma^{\,}_{5}\,
\gamma^{3},
\qquad
\psi^{u,d}\=:
\chi^{u,d},
\label{s1.9}
\end{equation}
and in terms of which
\begin{equation}
\begin{split}
\mathcal{L}=&\,
\bar{\chi}^{u}\,
\left(
\mathrm{i}\slash\!\!\!{\partial}
-
e\,\slash\!\!\!\!{A^{u}}
-
\eta
\right)\,
\chi^{u}
\\
&\,
+
\bar{\chi}^{d}\,
\left(
\mathrm{i}\slash\!\!\!{\partial}
-
e\,\slash\!\!\!\!{A^{d}}
+
\eta
\right)\,
\chi^{d}
\\
&\,
+
\lambda^{\,}_{\mathrm{R}}\,
\bar{\chi}^{u}\,
\left(
-\mathrm{i}
\Gamma^{1}
-
\Gamma^{2}
\right)\,
\chi^{d}
+
\lambda^{\,}_{\mathrm{R}}\,
\bar{\chi}^{d}\,
\left(
\mathrm{i}\Gamma^{1}
-
\Gamma^{2}
\right)\,
\chi^{u},
\end{split}
\label{s1.10}
\end{equation}
$\slash\!\!\!\!{A}\equiv\Gamma^{\mu}\,A^{\,}_{\mu}$, and
\begin{equation}
\Gamma^{\mu}\equiv
\gamma^{\,}_{5}\,\gamma^{3}\,\gamma^{\mu},
\qquad
\{\Gamma^{\mu},\Gamma^{\nu}\}=
2\,g^{\mu\nu},
\label{s1.11}
\end{equation}
\end{subequations}
for $\mu,\nu=0,1,2$. The signature of the Minkowski metric
is $g^{\,}_{\mu\nu}=\text{diag}(1,-1,-1)$.
Some useful properties of Dirac matrices are presented in Appendix
\ref{apendiceA}.

The Lagrangian density (\ref{s1.10}) is a special case of
\begin{subequations}
\begin{equation}
\begin{split}
\mathcal{L}=&\,
\bar\chi^{u}\,
\left(
\mathrm{i}
\slash\!\!\!{\partial}
-
\eta
\right)\,
\chi^{u}
+
\bar\chi^{d}
\left(
\mathrm{i}\slash\!\!\!{\partial}
+
\eta
\right)\,
\chi^{d}
-
e\,
\bar\chi^{u}\,
\slash\!\!\!\!{A^{u}}\,
\chi^{u}
\\
&\,
-
e\,
\bar\chi^{d}\,
\slash\!\!\!\!{A^{d}}\,
\chi^{d}
+
\lambda^{\,}_{\mathrm{R}}\,
\bar\chi^{u}\,
\slash\!\!\!{V}\,
\chi^{d}
+
\lambda^{\,}_{\mathrm{R}}\,
\bar\chi^{d}\,
\slash\!\!\!\!{W}\,
\chi^{u},
\end{split}
\label{1}
\end{equation}
with the choice
\begin{equation}
\begin{pmatrix}
V_0
\\
V_1
\\
V_2
\end{pmatrix}
\:=
\begin{pmatrix}
\;0
\\
-\mathrm{i}
\\
-1
\end{pmatrix},
\qquad
\begin{pmatrix}
W_0
\\
W_1
\\
W_2
\end{pmatrix}
\:=
\begin{pmatrix}
\;0
\\
+\mathrm{i}
\\
-1
\end{pmatrix}.
\end{equation}
\end{subequations}
We will leave the vectors $V^{\,}_{\mu}$ and $W^{\,}_{\mu}$
arbitrary throughout the perturbative calculations to come.
Notice that the Hermiticity condition for
the Lagrangian only demands that $W^{\ast}_{\mu}=V^{\,}_{\mu}$.

One fundamental property of the Lagrangian (\ref{1}) is its invariance
under reversal of time. The transformation law of the
Dirac fields under reversal of time is
\begin{equation}
\begin{split}
&
\chi^{u}\rightarrow
\mathrm{i}
\sigma^{\,}_{1}\otimes\tau^{\,}_{1}\,\chi^{d},
\qquad
\bar\chi^{u}\rightarrow
-\mathrm{i}
\bar\chi^{d}\,
\sigma^{\,}_{1}\otimes
\tau^{\,}_{1},
\\
&
\chi^{d}\rightarrow
-\mathrm{i}
\sigma^{\,}_{1}\otimes\tau^{\,}_{1}
\chi^{u},
\qquad
\bar\chi^{d}\rightarrow
+\mathrm{i}
\bar\chi^{u}\,
\sigma^{\,}_{1}\otimes\tau^{\,}_{1},
\label{2a}
\end{split}
\end{equation}
while the transformation law of the gauge fields is
\begin{equation}
A^{u,d}_{0}\rightarrow
+A^{d,u}_{0},
\qquad
A^{u,d}_{i}\rightarrow
-A^{d,u}_{i}.
\label{2b}
\end{equation}
Notice here the interchange between the flavors up and down.
The invariance of the Lagrangian density (\ref{1})
under reversal of time is achived if and only if
\begin{equation}
V^{\,}_{0}=W^{\,}_{0}=0.
\label{2c}
\end{equation}
Reversal of time does not restrict the spatial components $V^{\,}_{i}$
and $W^{\,}_{i}(=V^{*}_{i})$ for $i=1,2$.

Needed is the effective action generated for the fields
$A^{u}_{\mu}$
and
$A^{d}_{\mu}$
in the background
$V^{\,}_{\mu}$
and
$W^{\,}_{\mu}$
from integrating out the massive Dirac fermions. We are going to show that
\vskip 10 true pt
\begin{widetext}
\begin{equation}
\begin{split}
I[A^{u},A^{d}]=&\,
\int\frac{\mathrm{d}^{3}k}{(2\pi)^{3}}
\Bigg[
A^{u}_{\mu}(-k)
\left(
I^{(2)}_{uu}
\right)^{\mu\nu}(k)\,
A^{u}_{\nu}(k)
+
A^{u}_{\mu}(-k)\,
\left(
I^{(2)}_{ud}
\right)^{\mu\nu}(k)\,
A^{d}_{\nu}(k)
\\
&\,
+
A^{d}_{\mu}(-k)\,
\left(I^{(2)}_{du}\right)^{\mu\nu}(k)\,
A^{u}_{\nu}(k)
+
A^{d}_{\mu}(-k)\,
\left(I_{dd}^{(2)}\right)^{\mu\nu}(k)\,
A^{d}_{\nu}(k)
+
\cdots
\Bigg].
\label{3}
\end{split}
\end{equation}
\end{widetext}
The dots include terms of higher order than quadratic
in the gauge fields.
In this expression, $I^{(2)}_{IJ}$ represents the one-particle irreducible
(1PI) two-point vertex functions for the pair of gauge fields labeled by
$I=u,d$ and $J=u,d$.

\subsection{Propagators}

Our first task is to  choose
the propagators associated to the Lagrangian
(\ref{1}). To obtain the exact free propagators, we rewrite
the Lagrangian density (\ref{1}) as
\begin{equation}
\mathcal{L}^{0}\:=
\bar\chi^{I}\,
M^{IJ}\,
\chi^{J},
\label{37}
\end{equation}
with $I,J=u,d$. The matrix $M$ in momentum space is
\begin{equation}
M(p)=
\left(
\begin{array}{cc}
\displaystyle \slash\!\!\!p-\eta &
\displaystyle \lambda^{\,}_{\mathrm{R}}\slash\!\!\!V\\
\displaystyle  \lambda^{\,}_{\mathrm{R}}\,\slash\!\!\!\!W&
\displaystyle \slash\!\!\!p+\eta\\
\end{array}
\right).
\label{38}
\end{equation}
The propagator $S^{IJ}$ is defined to be
\begin{equation}
S(p)\:=
\mathrm{i}\,M^{-1}(p).
\end{equation}
The multiplicative factor $\mathrm{i}$ is chosen by convention.

Observe that,  aside from the propagators
$S^{uu}$ and $S^{dd}$, there are the mixed propagators $S^{ud}$ and
$S^{du}$. We define the inverse matrix
\begin{equation}
M^{-1}(p)=
\left(
\begin{array}{cc}
\displaystyle A &
\displaystyle B\\
\displaystyle  C&
\displaystyle D\\
\end{array}
\right),
\label{39}
\end{equation}
where $A,B,C$, and $D$ are matrices to be determined.
Imposing the condition $M M^{-1}=1$, we obtain the set of equations
\begin{subequations}
\begin{equation}
(\slash\!\!\!{p}-\eta)\,A
+
\lambda^{\,}_{\mathrm{R}}\,
\slash\!\!\!{V}\,C=1,
\label{40}
\end{equation}
\begin{equation}
(\slash\!\!\!{p}-\eta)\,B
+
\lambda^{\,}_{\mathrm{R}}\,
\slash\!\!\!{V}\,D=0,
\label{41}
\end{equation}
\begin{equation}
\lambda^{\,}_{\mathrm{R}}\,
\slash\!\!\!\!{W}\,A
+
(\slash\!\!\!{p}+\eta)\,C=0,
\label{42}
\end{equation}
and
\begin{equation}
\lambda^{\,}_{\mathrm{R}}\,
\slash\!\!\!\!{W}\,B
+
(\slash\!\!\!{p}+\eta)\,D=1.
\label{43}
\end{equation}
\end{subequations}
The formal solutions to these equations are
\begin{subequations}
\begin{equation}
S^{uu}(p)=
\frac{
\mathrm{i}
     }
     {
\slash\!\!\!{p}
-
\eta
-
\frac{
\lambda^{2}_{\mathrm{R}}}{p^{2}-\eta^{2}
     }\,
\slash\!\!\!{V}\,
(\slash\!\!\!{p}-\eta)\,
\slash\!\!\!\!{W}
     },
\label{44}
\end{equation}
\begin{equation}
S^{du}(p)=
-\lambda^{\,}_{\mathrm{R}}\,
\frac{
1
     }
     {
\slash\!\!\!{p}+\eta
     }\,
\slash\!\!\!\!{W}\,
\frac{
\mathrm{i}
     }
     {
\slash\!\!\!{p}
-
\eta
-
\frac{
\lambda^{2}_{\mathrm{R}}}{p^{2}-\eta^{2}
     }\,
\slash\!\!\!{V}\,
(\slash\!\!\!{p}-\eta)\,
\slash\!\!\!\!{W}
     },
\label{45}
\end{equation}
\begin{equation}
S^{dd}(p)=
\frac{
\mathrm{i}
     }
     {
\slash\!\!\!{p}
+
\eta
-
\frac{
\lambda^{2}_{\mathrm{R}}}{p^{2}-\eta^{2}
     }\,
\slash\!\!\!\!{W}\,
(\slash\!\!\!{p}+\eta)\,
\slash\!\!\!\!{V}
     },
\label{46}
\end{equation}
and
\begin{equation}
S^{ud}(p)=
-\lambda^{\,}_{\mathrm{R}}\,
\frac{
1
     }
     {
\slash\!\!\!{p}-\eta
     }\,
\slash\!\!\!{V}
\frac{
\mathrm{i}
     }
     {
\slash\!\!\!{p}
+
\eta
-
\frac{\lambda^{2}_{\mathrm{R}}}{p^{2}-\eta^{2}}\,
\slash\!\!\!\!{W}\,
(\slash\!\!\!{p}+\eta)\,
\slash\!\!\!{V}
     }.
\label{47}
\end{equation}
\end{subequations}
These propagators display an intricate matrix structure,
making diagrammatic calculations impracticable. Hence,
we will consider the limit
$|\lambda^{\,}_{\mathrm{R}}|\ll|\eta|$
and perform an expansion in powers of
${|\lambda^{\,}_{\mathrm{R}}|}/{|\eta|}$.

\section{Perturbative expansion in the Rashba spin-orbit coupling}
\label{sec: Perturbative Rashba Coupling}

\subsection{Feynman rules}

\begin{figure}[!t]
\centering
\includegraphics[scale=.65]{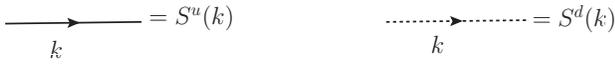}
\caption{Fermionic propagators.}
\label{f1}
\end{figure}

\begin{figure}[!t]
\centering
\includegraphics[scale=.5]{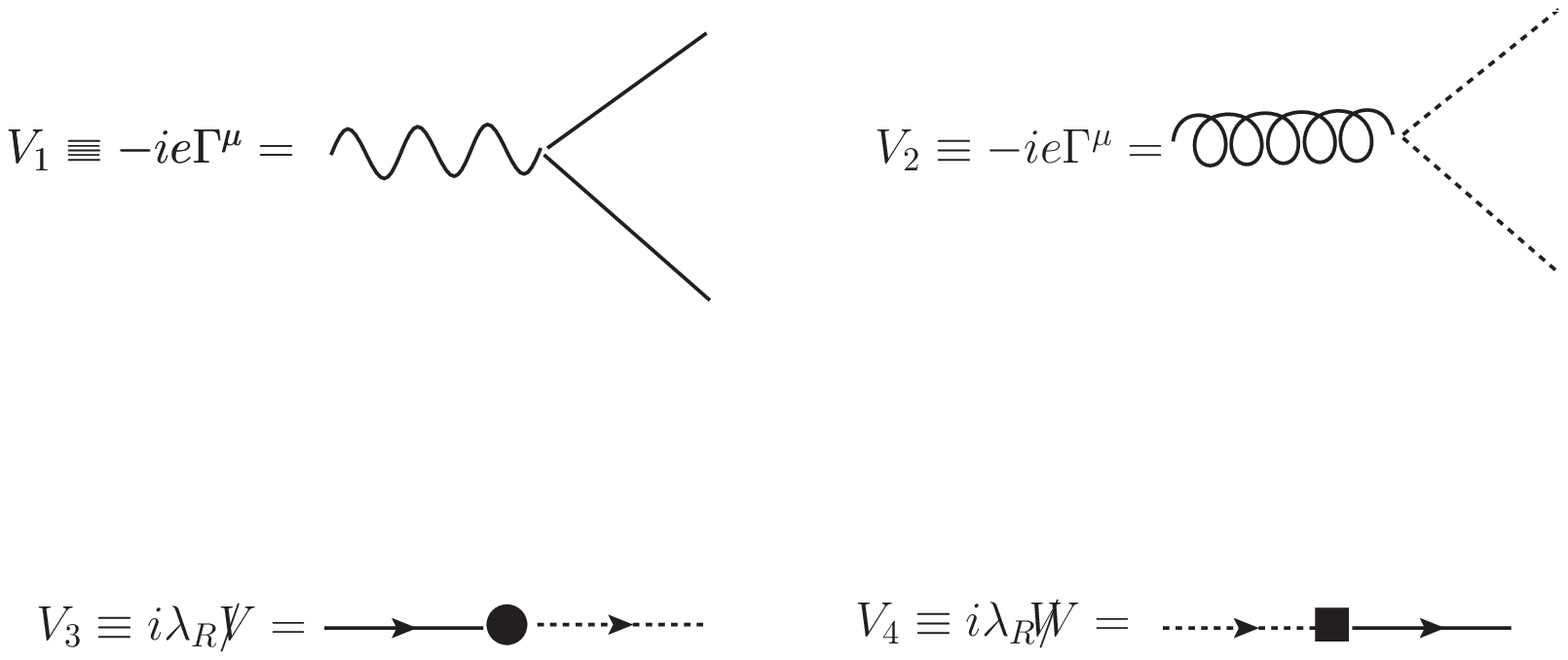}
\caption{
Vertices representing the coupling between the fermions and the gauge fields.
The wavy line in the vertex $V^{\,}_{1}$
refers to the gauge field $A^{u}_{\mu}$,
whereas the curly line in $V^{\,}_{2}$ refers to $A^{d}_{\mu}$.
        }
\label{f2}
\end{figure}

In order to develop perturbation theory, we use the following Feynman
rules associated to the Lagrangian density (\ref{1}).  The fermion
free propagators are defined by setting
$\lambda^{\,}_{\mathrm{R}}=0$ in Eqs. (\ref{44}-\ref{47}).
The non-vanishing propagators are
\begin{equation}
S^{u}(p)\:=
\frac{
\mathrm{i}
     }
     {
\slash\!\!\!{p}
-
\eta
+
\mathrm{i}\epsilon
     },
\qquad
S^{d}(p)\:=
\frac{
\mathrm{i}
     }
     {
\slash\!\!\!{p}
+
\eta
+
\mathrm{i}\epsilon
     },
\label{4}
\end{equation}
where we have introduced the
$\mathrm{i}\epsilon$ prescription to regulate poles.
These propagators are represented by the lines shown in
Fig.~\ref{f1}.
We have four types of vertices, as shown in Fig.~\ref{f2}.

\medskip

\subsection{Double Chern-Simons contributions}

\begin{figure}[!t]
\centering
\includegraphics[scale=.55]{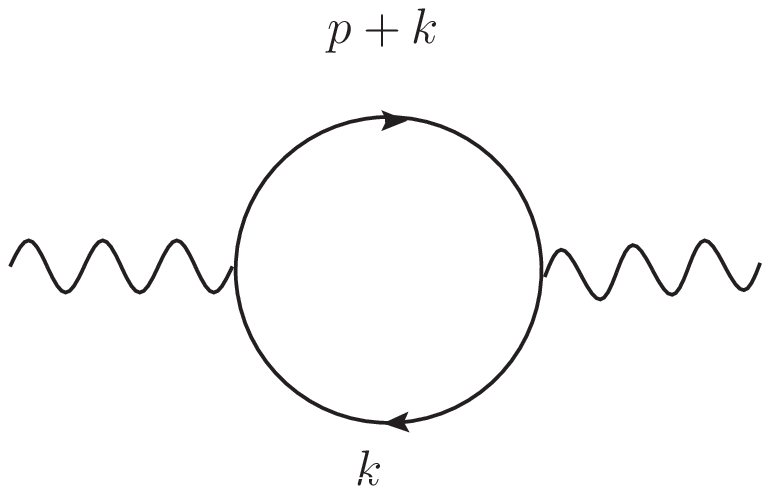}
\caption{Diagram contributing to the Chern-Simons term of $A^{u}_{\mu}$.}
\label{f3}
\end{figure}

\begin{figure}[!t]
\centering
\includegraphics[scale=.5]{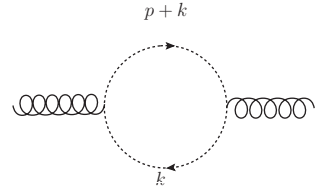}
\caption{Diagram contributing to the Chern-Simons term of $A^{d}_{\mu}$.}
\label{f4}
\end{figure}

We start by calculating the contributions of order
$e^{2}$ and $(e^{2}\,\lambda^{2}_{\mathrm{R}})^{0}$
to the 2-point 1PI vertex functions of the fields $A^{u}_{\mu}$ and
$A^{d}_{\mu}$, namely the ground-state expectation values
$\langle A^{u}_{\mu}\,A^{u}_{\nu}\rangle$
and
$\langle A^{d}_{\mu}\,A^{d}_{\nu}\rangle$.
These contributions are responsible for generating
the doubled Chern-Simons action
and correspond to the diagrams shown in
Figs.~\ref{f3} and \ref{f4}.  Up to this order, we do not have
contributions to the mixed ground-state expectation values
$\langle A^{u}_{\mu}\,A^{d}_{\nu}\rangle$ and
$\langle A^{d}_{\mu}\,A^{u}_{\nu}\rangle$.

According to the Feynman rules, the expression
corresponding to the  Feynman diagram~\ref{f3} is
\begin{equation}
\mathrm{i}(I^{(2)}_{uu})^{\mu\nu}(p)\Big{|}^{\,}_{e^{2}}=
e^{2}\,
\int\frac{\mathrm{d}^{3}k}{(2\pi)^3}
\text{Tr}\left[\Gamma^{\mu}\,S^{u}(p+k)\,\Gamma^{\nu}\,S^{u}(k)\right].
\label{5}
\end{equation}
This integral is linearly divergent but the Chern-Simons contribution
turns out to be finite.  Selecting only the Chern-Simons contribution,
we find
\begin{widetext}
\begin{equation}
\mathrm{i}(I^{(2)}_{uu})^{\mu\nu}(p)\Big{|}^{\,}_{e^{2}}=
-4\mathrm{i}e^{2}\,\eta\,
p^{\,}_{\alpha}\,
\epsilon^{\mu\alpha\nu}\,
\int\frac{\mathrm{d}^{3}k}{(2\pi)^{3}}\,
\frac{
1
     }
     {
[(p+k)^{2}-\eta^{2}+\mathrm{i}\epsilon][k^{2}-\eta^{2}+\mathrm{i}\epsilon]
     }
+
\cdots.
\label{6}
\end{equation}
\end{widetext}
For simplicity, from now on we will omit the $\mathrm{i}\epsilon$
prescription. The identity
\begin{equation}
\frac{1}{a\,b}=
\int\limits_{0}^{1}\mathrm{d}x\frac{1}{[a\,x+b\,(1-x)]^{2}},
\label{7}
\end{equation}
allows to perform the integration in the loop momentum.
One finds
\begin{widetext}
\begin{equation}
\mathrm{i}(I^{(2)}_{uu})^{\mu\nu}(p)\Big{|}^{\,}_{e^{2}}=
\frac{e^{2}}{2\pi}\eta\,
p^{\,}_{\alpha}\,
\epsilon^{\mu\alpha\nu}\,
\int\limits_{0}^{1}\mathrm{d}x\,
\frac{1}{[\eta^{2}-x\,(1-x)\,p^{2}]^{1/2}}
+
\cdots.
\label{8}
\end{equation}
\end{widetext}
For small momentum, i.e., $p^{2}/\eta^{2}\ll 1$, we perform
an expansion of the integrand in powers of $p^{2}/\eta^{2}$.
The result is the Chern-Simons kernel
\begin{equation}
\mathrm{i}(I^{(2)}_{uu})^{\mu\nu}(p)\Big{|}^{\,}_{e^{2}}=
+
\frac{e^{2}}{2\pi}\,
\frac{\eta}{|\eta|}\,
p^{\,}_{\alpha}\,
\epsilon^{\mu\alpha\nu}
+
\cdots.
\label{9}
\end{equation}
Higher order corrections in $p^{2}/\eta^{2}$ generate terms
quadratic in the gauge fields of higher order in the derivatives. For
example, there is the term
$\epsilon^{\mu\nu\rho}\,A^{\,}_{\mu}\Box\partial^{\,}_{\nu}\,A^{\,}_{\rho}$,
with the metric-dependent d'Alembert operator
$\Box\equiv\partial^{\,}_{\mu}\partial^{\mu}$.
This term is infra-red irrelevant by power counting.

Similarly, the contribution of the Feynman diagram
depicted in Fig.~\ref{f4}
to the 2-point 1PI vertex function of the gauge field $A^{d}_{\mu}$ is
\begin{equation}
\mathrm{i}(I^{(2)}_{dd})^{\mu\nu}(p)\Big{|}^{\,}_{e^{2}}=
e^{2}\,
\int\frac{\mathrm{d}^{3}k}{(2\pi)^{3}}\,
\text{Tr}\left[\Gamma^{\mu}\,S^{d}(p+k)\,\Gamma^{\nu}\,S^{d}(k)\right].
\label{10}
\end{equation}
The only difference between the right-hand side of
Eqs.~(\ref{10}) and (\ref{9}) is the sign with which the
mass $\eta$ enters the free propagators. Hence, we find
\begin{equation}
\mathrm{i}(I^{(2)}_{dd})^{\mu\nu}(p)\Big{|}^{\,}_{e^{2}}=
-
\frac{e^{2}}{2\pi}\,
\frac{\eta}{|\eta|}\,
p^{\,}_{\alpha}\,
\epsilon^{\mu\alpha\nu}
+
\cdots.
\label{11}
\end{equation}

Collecting Eqs.~(\ref{9}) and (\ref{11}), we find for
the effective field theory the double Chern-Simons theory
\begin{equation}
\mathcal{L}^{\,}_{\mathrm{eff}}\Big{|}^{\,}_{e^{2}}=
\frac{e^{2}}{2\pi}\,
\frac{\eta}{|\eta|}\,
\epsilon^{\mu\alpha\nu}
\left(
A^{u}_{\mu}\partial^{\,}_{\alpha}A^{u}_{\nu}
-
A^{d}_{\mu}\partial^{\,}_{\alpha}A^{d}_{\nu}
\right)
+
\cdots
\label{12}
\end{equation}
up to order $e^{2}$ and $(e^{2}\,\lambda^{2}_{\mathrm{R}})^{0}$
in the couplings and to the first non-vanishing order in a gradient
expansion.

\begin{figure}[!t]
\centering
\includegraphics[scale=.45]{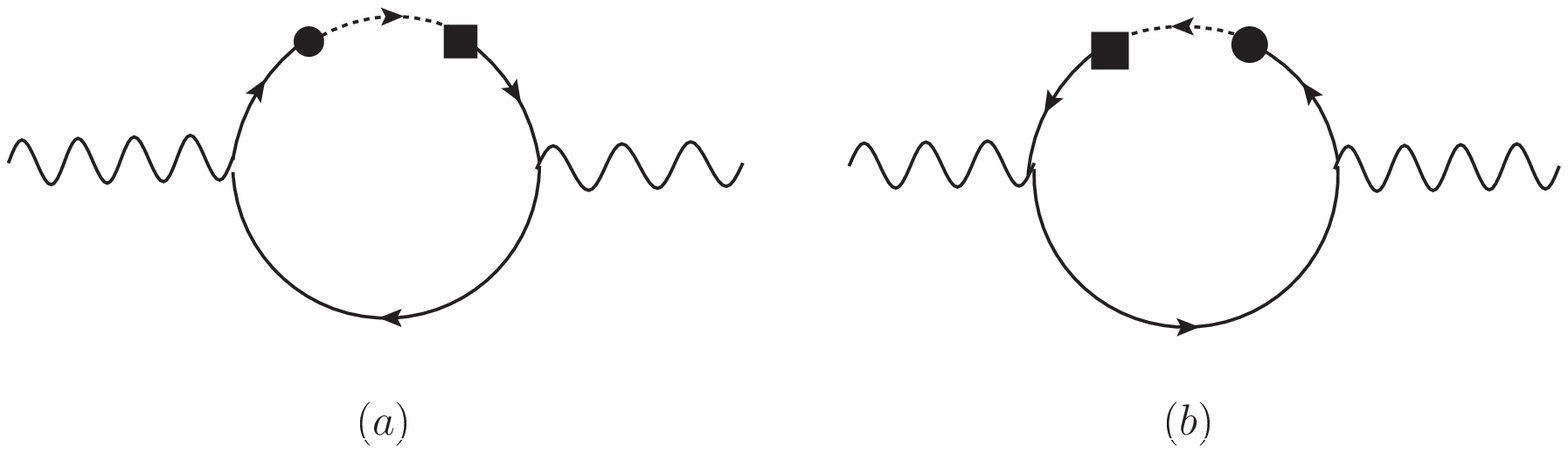}
\caption{
Rashba contributions of order $e^{2}\,\lambda^{2}_{\mathrm{R}}$
to $\langle A^{u}_{\mu}\,A^{u}_{\nu}\rangle$.
        }
\label{f5}
\end{figure}

\begin{figure}[!t]
\centering
\includegraphics[scale=.5]{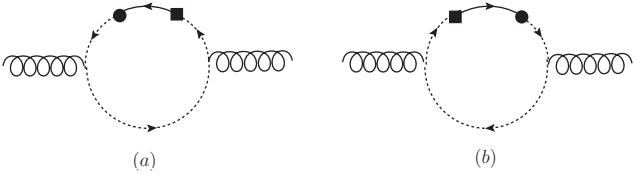}
\caption{
Rashba contributions of order
$e^{2}\,\lambda^{2}_{\mathrm{R}}$
to
$\langle A^{d}_{\mu}\,A^{d}_{\nu}\rangle$.
        }
\label{f6}
\end{figure}

\subsection{Rashba corrections}

We are after the corrections of order
$e^{2}\,\lambda^{2}_{\mathrm{R}}$ to
the 2-point functions of $A^{u}_{\mu}$ and $A^{d}_{\mu}$.

\subsubsection{The 1PI vertex function $\langle A^{u}_{\mu}A^{u}_{\nu}\rangle$}

Starting with the function $\langle A^{u}_{\mu}\,A^{u}_{\nu}\rangle$,
we have the contributions of the two diagrams of Fig.~\ref{f5},
with the corresponding expressions
\begin{widetext}
\begin{equation}
\begin{split}
\mathrm{i}(I^{(2)}_{uu})^{\mu\nu}(p)\,
\Big{|}^{\,}_{\lambda^{2}_{\mathrm{R}}\,e^{2}}=&\,
-
\lambda^{2}_{\mathrm{R}}\,
e^{2}\,
\int\frac{\mathrm{d}^{3}k}{(2\pi)^{3}}\,
\text{Tr}
\left[
\Gamma^{\mu}\,
S^{u}(p+k)\,
\slash\!\!\!{V}\,
S^{d}(p+k)\,
\slash\!\!\!\!{W}\,
S^{u}(p+k)\,
\Gamma^{\nu}\,
S^{u}(k)
\right]
\\
&\,
-\lambda^{2}_{\mathrm{R}}\,
e^{2}\,
\int\frac{\mathrm{d}^{3}k}{(2\pi)^{3}}\,
\text{Tr}
\left[
\Gamma^{\mu}\,
S^{u}(p+k)\,
\Gamma^{\nu}\,
S^{u}(k)\,
\slash\!\!\!{V}\,
S^{d}(k)\,\,\slash\!\!\!\!{W}\,
S^{u}(k)
\right].
\label{13}
\end{split}
\end{equation}
These integrals are finite. As the leading terms do not depend on the
external momentum, we can simplify the calculation by setting $p=0$,
\begin{equation}
\begin{split}
\mathrm{i}(I^{(2)}_{uu})^{\mu\nu}(p=0)\,
\Big{|}^{\,}_{\lambda^{2}_{\mathrm{R}}\,e^{2}}=&\,
-\lambda^{2}_{\mathrm{R}}\,
e^{2}\,
\int\frac{\mathrm{d}^{3}k}{(2\pi)^{3}}\,
\text{Tr}
\left[
\Gamma^{\mu}\,
S^{u}(k)\,
\slash\!\!\!{V}\,
S^{d}(k)\,\,
\slash\!\!\!\!{W}\,
S^{u}(k)\,
\Gamma^{\nu}\,
S^{u}(k)
\right]
\\
&\,
-
\lambda^{2}_{\mathrm{R}}\,
e^{2}\,
\int\frac{\mathrm{d}^{3}k}{(2\pi)^{3}}\,
\text{Tr}
\left[
\Gamma^{\nu}\,
S^{u}(k)\,
\slash\!\!\!{V}\,
S^{d}(k)\,\,
\slash\!\!\!\!{W}\,
S^{u}(k)\,
\Gamma^{\mu}\,
S^{u}(k)
\right],
\label{14}
\end{split}
\end{equation}
\end{widetext}
where we used the cyclicity of the trace to rewrite the second
term in such way that it differs from the first one only by the change
$\mu\leftrightarrow \nu$.  The calculation of this expression
is simplified by the introduction of the rank-4 tensor
\begin{equation}
\begin{split}
A^{\mu\nu\alpha\beta}\:=&\,
\int\frac{\mathrm{d}^{3}k}{(2\pi)^{3}}\,
\text{Tr}
\Big[
\Gamma^{\mu}\,
S^{u}(k)\,
\Gamma^{\alpha}\,
S^{d}(k)
\\
&\,
\times
\Gamma^{\beta}\,
S^{u}(k)\,
\Gamma^{\nu}\,
S^{u}(k)
\Big].
\end{split}
\end{equation}
A detailed evaluation of the trace as well as the loop integral is
done in Appendix \ref{apendiceB}, where it is shown that
\begin{equation}
A^{\mu\nu\alpha\beta}=
\frac{\mathrm{i}}{12\,\pi\,|\eta|}\,
\left(
g^{\mu\nu}\,
g^{\alpha\beta}
-
g^{\mu\alpha}\,
g^{\nu\beta}
-
g^{\mu\beta}\,
g^{\nu\alpha}
\right).
\label{22}
\end{equation}
Notice that
$A^{\mu\nu\alpha\beta}$ is symmetric in the exchange of the pairs
of indices $(\mu,\nu)$ and $(\alpha,\beta)$. Equation
(\ref{14}) becomes
\begin{widetext}
\begin{equation}
\mathrm{i}(I^{(2)}_{uu})^{\mu\nu}(p=0)\Big{|}^{\,}_{\lambda^{2}_{\mathrm{R}}\,e^{2}}=
-
\frac{\mathrm{i}\lambda^{2}_{\mathrm{R}}\,e^{2}}{6\,\pi\,|\eta|}
\left[g^{\mu\nu}\,(V\cdot W)-V^{\mu}\,W^{\nu}-V^{\nu}\,W^{\mu}\right].
\label{23}
\end{equation}
The effective Lagrangian in the coordinate space reads
\begin{equation}
\mathcal{L}^{uu}_{\mathrm{eff}}\Big{|}^{\,}_{\lambda^{2}_{\mathrm{R}}\,e^{2}}=
-
\frac{\lambda^{2}_{\mathrm{R}}\,e^{2}}{6\,\pi\,|\eta|}
\left[
(A^{u}\cdot A^{u})\,
(V\cdot W)
-
2\,
(A^{u}\cdot V)\,
(A^{u}\cdot W)
\right].
\label{24}
\end{equation}
\end{widetext}

\subsubsection{The 1PI vertex function
$\langle A^{d}_{\mu}\,A^{d}_{\nu}\rangle$}

We turn our attention to the 1PI vertex function
$\langle A^{d}_{\mu}\,A^{d}_{\nu}\rangle$.
The one-loop contribution is given by the Feynman
diagrams of Fig. \ref{f6},
\begin{widetext}
\begin{equation}
\begin{split}
\mathrm{i}(I^{(2)}_{dd})^{\mu\nu}(p)\Big{|}^{\,}_{\lambda^{2}_{\mathrm{R}}\,e^{2}}=&\,
-\lambda^{2}_{\mathrm{R}}\,
e^{2}\,
\int\frac{\mathrm{d}^{3}k}{(2\pi)^{3}}\,
\text{Tr}
\left[
\Gamma^{\mu}\,
S^{d}(p+k)\,
\Gamma^{\nu}\,
S^{d}(k)\,\,
\slash\!\!\!\!{W}\,
S^{u}(k)\,
\slash\!\!\!{V}\,
S^{u}(k)
\right]
\\
&\,
-
\lambda^{2}_{\mathrm{R}}\,
e^{2}
\int\frac{\mathrm{d}^{3}k}{(2\pi)^{3}}\,
\text{Tr}
\left[
\Gamma^{\mu}\,
S^{d}(p+k)\,\,
\slash\!\!\!\!{W}\,
S^{u}(p+k)\,
\slash\!\!\!{V}\,
S^{d}(p+k)\,
\Gamma^{\nu}\,
S^{d}(k)
\right].
\end{split}
\label{25}
\end{equation}
Equation (\ref{13}) is mapped into (\ref{25}) by changing
the sign of the mass, $\eta\rightarrow -\eta$, and interchanging
$\slash\!\!\!\!{V}$ and $\slash\!\!\!\!{W}$.
Hence,
\begin{equation}
\mathrm{i}(I^{(2)}_{dd})^{\mu\nu}(p=0)\Big{|}^{\,}_{\lambda^{2}_{\mathrm{R}}\,e^{2}}=
-\frac{\mathrm{i}\lambda^{2}_{\mathrm{R}}\,e^{2}}{6\,\pi\,|\eta|}
\left[
g^{\mu\nu}\,
(V\cdot W)
-
V^{\mu}\,
W^{\nu}
-
V^{\nu}\,
W^{\mu}
\right]
\label{26}
\end{equation}
and the corresponding effective Lagrangian density in the coordinate space
\begin{equation}
\mathcal{L}^{dd}_{\mathrm{eff}}
\Big{|}^{\,}_{\lambda^{2}_{\mathrm{R}}\,e^{2}}=
-\frac{\lambda_{\mathrm{R}}^{2}\,e^{2}}{6\,\pi\,|\eta|}\,
\left[
(A^{d}\cdot A^{d})\,
(V\cdot W)
-
2\,
(A^{d}\cdot V)\,
(A^{d}\cdot W)
\right].
\label{27}
\end{equation}
\end{widetext}

\subsubsection{The 1PI vertex function
$\langle A^{u}_{\mu}\,A^{d}_{\nu}\rangle$}

\begin{figure}[!t]
\centering
\includegraphics[scale=.5]{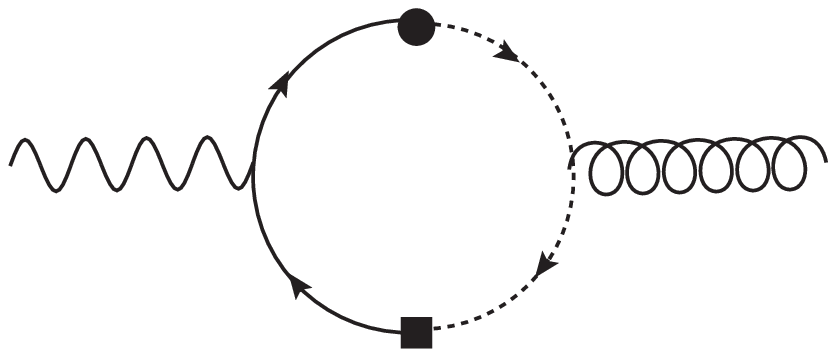}
\caption{
Rashba contributions of order
$e^{2}\,\lambda^{2}_{\mathrm{R}}$
to
$\langle A^{u}_{\mu}\,A^{d}_{\nu}\rangle$.
        }
\label{f7}
\end{figure}

The 1PI vertex function
$\langle A^{u}_{\mu}\,A^{d}_{\nu}\rangle$
has the one-loop contribution shown in the
Feynman diagram of Fig.~\ref{f7},
\begin{widetext}
\begin{equation}
\mathrm{i}(I^{(2)}_{ud})^{\mu\nu}(p)\Big{|}^{\,}_{\lambda^{2}_{\mathrm{R}}\,e^{2}}=
-
\lambda^{2}_{\mathrm{R}}\,e^{2}
\int\frac{\mathrm{d}^{3}k}{(2\pi)^{3}}\,
\text{Tr}
\left[
\Gamma^{\mu}\,
S^{u}(p+k)\,
\slash\!\!\!{V}\,
S^{d}(p+k)\,
\Gamma^{\nu}\,
S^{d}(k)\,\,
\slash\!\!\!\!{W}\,
S^{u}(k)
\right].
\label{28}
\end{equation}
The trace and loop integral are performed in Appendix
\ref{apendiceB}. The result for $p=0$ is
\begin{equation}
\mathrm{i}(I^{(2)}_{ud})^{\mu\nu}(p)\Big{|}^{\,}_{\lambda^{2}_{\mathrm{R}}\,e^{2}}=
\frac{\mathrm{i}\lambda^{2}_{\mathrm{R}}\,e^{2}}{6\,\pi\,|\eta|}\,
\left[
g^{\mu\nu}\,
(V\cdot W)
-
V^{\mu}\,
W^{\nu}
-
V^{\nu}\,
W^{\mu}
\right].
\label{29}
\end{equation}
By turning this equation to the coordinate space, it follows that
\begin{equation}
\mathcal{L}^{ud}_{\mathrm{eff}}\Big{|}^{\,}_{\lambda^{2}_{\mathrm{R}}\,e^{2}}=
\frac{\lambda^{2}_{\mathrm{R}}\,e^{2}}{6\,\pi\,|\eta|}\,
\left[
(A^{u}\cdot A^{d})\,
(V\cdot W)
-
(A^{u}\cdot V)\,
(A^{d}\cdot W)
-
(A^{u}\cdot W)\,
(A^{d}\cdot V)
\right].
\label{30}
\end{equation}
\end{widetext}

\subsubsection{The 1PI vertex function
$\langle A^{d}_{\mu}\,A^{u}_{\nu}\rangle$}

\begin{figure}[!t]
\centering
\includegraphics[scale=.5]{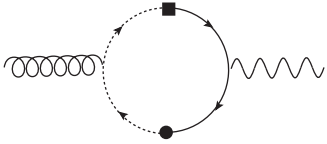}
\caption{
Rashba contributions of order
$\lambda^{2}_{\mathrm{R}}\,e^{2}$
to
$\langle A^{d}_{\mu}\,A^{u}_{\nu}\rangle$.
        }
\label{f8}
\end{figure}

The 1PI vertex function $\langle A^{d}_{\mu}\,A^{u}_{\nu}\rangle$
has the one-loop contribution shown in the Feynman diagram
of Fig.~\ref{f8},
\begin{widetext}
\begin{equation}
\mathrm{i}(I^{(2)}_{du})^{\mu\nu}(p)\Big{|}^{\,}_{\lambda^{2}_{\mathrm{R}}\,e^{2}}=
-\lambda^{2}_{\mathrm{R}}\,e^{2}\,
\int\frac{\mathrm{d}^{3}k}{(2\pi)^{3}}\,
\text{Tr}
\left[
\Gamma^{\mu}\,
S^{d}(p+k)\,\,
\slash\!\!\!\!{W}\,
S^{u}(p+k)\,
\Gamma^{\nu}\,
S^{u}(k)\,
\slash\!\!\!{V}\,
S^{d}(k)
\right].
\label{31}
\end{equation}
Equation (\ref{28}) maps into Eq.\ (\ref{31})
by doing the substitutions
$V\leftrightarrow W$ and $\eta\rightarrow -\eta$.
Under these substitutions, Eq.\ (\ref{29}) is turned into
\begin{equation}
\mathrm{i}(I^{(2)}_{du})^{\mu\nu}(p)\Big{|}^{\,}_{\lambda^{2}_{\mathrm{R}}\,e^{2}}=
\frac{\mathrm{i}\lambda^{2}_{\mathrm{R}}\,e^{2}}{6\,\pi\,|\eta|}\,
\left[
g^{\mu\nu}\,
(V\cdot W)
-
V^{\mu}\,
W^{\nu}
-
V^{\nu}\,
W^{\mu}
\right].
\label{32}
\end{equation}
The effective Lagrangian density reads
\begin{equation}
\mathcal{L}^{du}_{\mathrm{eff}}\Big{|}^{\,}_{\lambda^{2}_{\mathrm{R}}\,e^{2}}=
\frac{\lambda^{2}_{\mathrm{R}}\,e^{2}}{6\,\pi\,|\eta|}\,
\left[
(A^{u}\cdot A^{d})\,
(V\cdot W)
-
(A^{u}\cdot V)\,
(A^{d}\cdot W)
-
(A^{u}\cdot W)\,
(A^{d}\cdot V)
\right].
\label{33}
\end{equation}

\subsubsection{Summary}

Adding Eqs.\ (\ref{24}), (\ref{27}), (\ref{30}), and (\ref{33}),
we obtain
\begin{subequations}
\begin{align}
\mathcal{L}^{\,}_{\mathrm{eff}}\Big{|}^{\,}_{\lambda^{2}_{\mathrm{R}}\,e^{2}}=&\,
\mathcal{L}^{uu}_{\mathrm{eff}}\Big{|}^{\,}_{\lambda^{2}_{\mathrm{R}}\,e^{2}}
+
\mathcal{L}^{dd}_{\mathrm{eff}}\Big{|}^{\,}_{\lambda^{2}_{\mathrm{R}}\,e^{2}}
+
\mathcal{L}^{ud}_{\mathrm{eff}}\Big{|}^{\,}_{\lambda^{2}_{\mathrm{R}}\,e^{2}}
+
\mathcal{L}^{du}_{\mathrm{eff}}\Big{|}^{\,}_{\lambda^{2}_{\mathrm{R}}\,e^{2}}
\nonumber\\
=&\,
-\frac{e^{2}\lambda^{2}_{\mathrm{R}}}{3\,\pi\,|\eta|}\,
\left[
\frac{1}{2}\,
(A^{(-)})^{2}\,
(V\cdot W)
-
[(A^{(-)}\cdot V]\,
[A^{(-)}\cdot W]
\right],
\label{34}
\end{align}
where we defined the gauge fields
\begin{equation}
A^{(\pm)}_{\mu}\equiv A^{u}_{\mu}\pm A^{d}_{\mu}.
\end{equation}
\end{subequations}
Thus, with the condition (\ref{2c}), the time-reversal-symmetric
one-loop effective action is
\begin{subequations}
\begin{align}
I[A^{(+)},A^{(-)}]=&\,
\frac{e^{2}}{4\pi}\,
\int\mathrm{d}^{3}x\,
\Bigg\{
\frac{\eta}{|\eta|}\,
\epsilon^{\mu\alpha\nu}\,
\left(
A^{(+)}_{\mu}\partial^{\,}_{\alpha}A^{(-)}_{\nu}
+
A^{(-)}_{\mu}\partial^{\,}_{\alpha}A^{(+)}_{\nu}
\right)
\nonumber\\
&\,
+
\frac{2\,\lambda^{2}_{\mathrm{R}}}{3\,|\eta|}\,
\left[
(A^{(-)}_{0})^{2}\,
(\vec{V}\cdot\vec{W})
+
A^{(-)}_{i}\,
A^{(-)}_{j}\,
\left(
\delta^{\,}_{ij}\,
\vec{V}\cdot\vec{W}
-
V^{\,}_{i}\,W_j
-
V^{\,}_{j}\,
W^{\,}_{i}
\right)
\right]
\Bigg\},
\label{cea}
\end{align}
where
\begin{equation}
\vec{V}\cdot\vec{W}\equiv
V^{\,}_{i}\,
W^{\,}_{i}.
\end{equation}
\end{subequations}
For the particular case of the Rashba spin-orbit coupling,
$V^{\top}=(0,-\mathrm{i},-1)$
and
$W^{\top}=(0,+\mathrm{i},-1)$
and this effective action reduces to
\begin{equation}
I[A^{(+)},A^{(-)}]=
\frac{e^{2}}{4\pi}\,
\int\mathrm{d}^{3}x
\left[
\frac{\eta}{|\eta|}\,
\epsilon^{\mu\alpha\nu}\,
\left(
A^{(+)}_{\mu}\partial^{\,}_{\alpha}A^{(-)}_{\nu}
+
A^{(-)}_{\mu}\partial^{\,}_{\alpha}A^{(+)}_{\nu}
\right)
+
\frac{4\,\lambda^{2}_{\mathrm{R}}}{3\,|\eta|}\,
\left(A^{(-)}_{0}\right)^{2}
\right].
\label{rea}
\end{equation}
\end{widetext}
The coupling constant $e^{2}$ multiplies the integrand in this effective
action.
Hence, it can be absorbed by the rescaling
\begin{equation}
|e|\, A^{(+)}_{\mu}\to
A^{(+)}_{\mu},
\qquad
|e|\, A^{(-)}_{\mu}\to
A^{(-)}_{\mu},
\end{equation}
of the gauge fields.
For convenience, we also do the redefinition
\begin{equation}
\frac{4}{3}\,\lambda^{2}_{\mathrm{R}}\to
\lambda^{2}_{\mathrm{R}}
\end{equation}
of the Rashba spin-orbit coupling.
Finally, we choose without loss of generality the sign
\begin{equation}
\mathrm{sgn}(\eta)=-
\end{equation}
for the intrinsic spin-orbit coupling.
In this way, we arrive at Eq.%
~(\ref{eq: first main result of the paper}).

\section{Edge Theory}
\label{sec: Edge Theory}

This section is devoted to deriving the bulk-edge correspondence when
the effective action~(\ref{eq: first main result of the paper})
is defined on a manifold with boundaries. To this end, we need
to extract from the effective action%
~(\ref{eq: first main result of the paper})
the effective Lagrangian density
\begin{widetext}
\begin{equation}
\mathcal{L}^{\,}_{\mathrm{eff}}=
-\frac{1}{4\pi}\,
\left[
\epsilon^{\mu\alpha\nu}\,
\left(
A^{(+)}_{\mu}\,
\partial^{\,}_{\alpha}
A^{(-)}_{\nu}
+
A^{(-)}_{\mu}\,
\partial^{\,}_{\alpha}A^{(+)}_{\nu}
\right)
-
\frac{\lambda^{2}_{\mathrm{R}}}{|\eta|}\,
\left(A^{(-)}_{0}\right)^{2}
\right].
\label{36a}
\end{equation}
\end{widetext}

\subsection{Pure Chern-Simons Theory}
\label{subsec: Pure Chern-Simons Theory}

As a warm up we first study
the bulk-edge correspondence in the absence of Rashba spin-orbit coupling.
In doing so, we shall emphasize the
ingredients that will be useful for the extension to the case
with Rashba spin-orbit coupling.

The effective action
(\ref{eq: first main result of the paper})
reduces to the double Chern-Simons action
\begin{equation}
S^{\,}_{\mathrm{CS}}=
-\frac{1}{4\pi}\,
\int\mathrm{d}^{3}x\,
\epsilon^{\mu\nu\lambda}\,
\left(
A^{(+)}_{\mu}\partial^{\,}_{\nu}A^{(-)}_{\lambda}
+
A^{(-)}_{\mu}\partial^{\,}_{\nu}A^{(+)}_{\lambda}
\right)
\label{1.1}
\end{equation}
when $\lambda^{\,}_{\mathrm{R}}=0$.
The variation of the Lagrangian density
(\ref{36a}) with $\lambda^{\,}_{\mathrm{R}}=0$
under the gauge transformations
\begin{equation}
A^{(+)}_{\mu}\rightarrow
A^{(+)}_{\mu}
+
\partial^{\,}_{\mu}\Lambda^{(+)},
\qquad
A^{(-)}_{\mu}\rightarrow
A^{(-)}_{\mu}
+
\partial^{\,}_{\mu}\Lambda^{(-)},
\label{1.2}
\end{equation}
is a total derivative. Thus, if the manifold has no boundary,
the theory is gauge invariant. In this case, we have the
freedom to fix any one of the components of $A^{(+)}_{\mu}$
and any one of the components of $A^{(-)}_{\mu}$. Correspondingly,
the equations of motion
\begin{equation}
0=
\frac{\delta S^{\,}_{\mathrm{CS}}}{\delta A^{(+)}_{\mu}}=
\epsilon^{\mu\nu\lambda}\,
\partial^{\,}_{\nu}A^{(-)}_{\lambda}
\label{1.6}
\end{equation}
and
\begin{equation}
0=
\frac{\delta S^{\,}_{\mathrm{CS}}}{\delta A^{(-)}_{\mu}}=
\epsilon^{\mu\nu\lambda}\,
\partial^{\,}_{\nu}A^{(+)}_{\lambda}
\label{1.7}
\end{equation}
dictate that
\begin{equation}
F^{(-)}_{\mu\nu}\:=
\partial^{\,}_{\mu}A^{(-)}_{\nu}
-
\partial^{\,}_{\nu}A^{(-)}_{\mu}=0,
\end{equation}
and
\begin{equation}
F^{(+)}_{\mu\nu}\:=
\partial^{\,}_{\mu}A^{(+)}_{\nu}
-
\partial^{\,}_{\nu}A^{(+)}_{\mu}=0,
\end{equation}
respectively.
Hence, the doubled Chern-Simons action does not support gapless excitations
when two-dimensional space has no boundary.

This freedom to fix all the components of the gauge fields
is lost if the space manifold has a boundary.
To appreciate this point, we choose a manifold $\Omega$ in two-dimensional
position space with an edge
running along the $x$ axis at $y=0$, i.e.,
\begin{equation}
\Omega\:=\{(x,y)|x\in \mathbb{R},y\le 0\},
\label{eq: def space only manifold}
\end{equation}
in addition to the time coordinate defined by $-\infty<t<+\infty$.
The 3-dimensional manifold over which the
doubled Chern-Simons Lagrangian density
\begin{equation}
\mathcal{L}^{\,}_{\mathrm{CS}}=
-\frac{1}{4\pi}\,
\epsilon^{\mu\nu\lambda}\,
\left(
A^{(+)}_{\mu}\partial^{\,}_{\nu}A^{(-)}_{\lambda}
+
A^{(-)}_{\mu}\partial^{\,}_{\nu}A^{(+)}_{\lambda}
\right)
\label{eq: def double CS Lagrangian}
\end{equation}
is to be integrated is thus
\begin{equation}
\Omega\times\mathbb{R}.
\label{eq: def space time manifold}
\end{equation}
Under the gauge transformations (\ref{1.2}),
the variation of the doubled Chern-Simons action
(\ref{1.1})
is the edge action
\begin{equation}
\begin{split}
\delta S^{\,}_{\mathrm{CS}}=&\,
-\frac{1}{4\pi}\,
\int\mathrm{d}x\mathrm{d}t\,
\Lambda^{(+)}\,
\left(
\partial^{\,}_{t}A_1^{(-)}
-
\partial^{\,}_{x}A^{(-)}_{0}
\right)
\Big{|}_{y=0}
\\
&\,
-\frac{1}{4\pi}\,
\int\mathrm{d}x\,\mathrm{d}t\,
\Lambda^{(-)}\,
\left(
\partial^{\,}_{t}A^{(+)}_{1}
-
\partial^{\,}_{x}A^{(+)} _{0}
\right)
\Big{|}^{\,}_{y=0}.
\label{1.3}
\end{split}
\end{equation}
Evidently, gauge invariance is lost for arbitrary
$\Lambda^{(+)}$
and
$\Lambda^{(-)}$
and so is the freedom to fix all the components of the gauge fields.

\subsubsection{Restoring gauge symmetry by restricting
the allowed gauge transformations}

One way to preserve the gauge invariance
on the space manifold (\ref{eq: def space only manifold})
and the space-time manifold (\ref{eq: def space time manifold})
is to restrict the gauge transformation in Eq.%
~(\ref{1.2})
by imposing the conditions
\begin{equation}
\Lambda^{(+)}\Big{|}_{y=0}=\Lambda^{(-)}\Big{|}_{y=0}=0
\label{1.4}
\end{equation}
for any coordinate $x$ along the edge and any time $t$.
Restricting the allowed functions $\Lambda^{(\pm)}$ by
imposing the constraint (\ref{1.4}) on the edge restores gauge invariance.
However, this gauge symmetry, restricted as it is on the edge,
allows for gapless degrees of freedom
to be supported on the boundary, as we demonstrate now.

We fix the gauge fields $A^{(+)}_{0}$ and $A^{(-)}_{0}$ by demanding that
they be proportional to the gauge fields $A^{(-)}_{1}$ and $A^{(+)}_{1}$,
respectively,
\begin{equation}
A^{(+)}_{0}=
v\,A^{(-)}_{1},
\qquad
A^{(-)}_{0}=
v\,A^{(+)}_{1}.
\label{1.5}
\end{equation}
The proportionality constant $v$ is arbitrary and carries
the dimension of velocity. It will shortly be identified
with the characteristic velocity of the edge states. The
arbitrariness in choosing $v$ reflects the fact that
the value of $v$ is fixed by the contributions to the effective action
of higher order in the derivative expansion than the
leading terms that have been kept, i.e., $v$ is
independent of the microscopic physics encoded
by the double Chern-Simons action.

As the components $A^{(+)}_{0}$ and $A^{(-)}_{0}$
are not independent dynamical degrees of freedom
anymore, their equations of motion,
Eqs.~(\ref{1.7}) and (\ref{1.6}) with $\mu=0$, respectively,
become the constraints
\begin{equation}
F^{(-)}_{12}\:=
\partial^{\,}_{x}A^{(-)}_{2}
-
\partial^{\,}_{y}A^{(-)}_{1}=0
\label{1.6 bis}
\end{equation}
and
\begin{equation}
F^{(+)}_{12}\:=
\partial^{\,}_{x}A^{(+)}_{2}
-
\partial^{\,}_{y}A^{(+)}_{1}=0,
\label{1.7 bis}
\end{equation}
respectively. Both constraints are met by
\begin{equation}
A^{(-)}_{i}=\partial^{\,}_{i}\varphi^{(-)},
\qquad
A^{(+)}_{i}=\partial^{\,}_{i}\varphi^{(+)},
\label{1.8}
\end{equation}
for $i=x,y$ if the scalar fields
$\varphi^{(-)}$
and
$\varphi^{(+)}$
are smooth.

One verifies that the action (\ref{1.1})
becomes
\begin{widetext}
\begin{equation}
S^{\,}_{\mathrm{edge}}=
\frac{1}{4\pi}\,
\int\mathrm{d}x\,\mathrm{d}t\,
\left(
\partial^{\,}_{t}\varphi^{(-)}\partial^{\,}_{x}\varphi^{(+)}
+
\partial^{\,}_{t}\varphi^{(+)}\partial^{\,}_{x}\varphi^{(-)}
-
v\,
\partial^{\,}_{x}\varphi^{(+)}\partial^{\,}_{x}\varphi^{(+)}
-
v\,
\partial^{\,}_{x}\varphi^{(-)}\partial^{\,}_{x}\varphi^{(-)}
\right)
\label{1.9}
\end{equation}
\end{widetext}
if we make use of Eqs.\
(\ref{1.5})
and
(\ref{1.8}).
This is the action for a pair of massless relativistic
counter propagating chiral bosonic modes in (1+1)-dimensional space and time.
As promised, gapless excitations are supported by the edge even though the
theory in the bulk is massive.

\subsubsection{Restoring gauge symmetry by adding dynamical degrees of freedom on the edges}
\label{subsec: Restoring gauge symmetry by adding ... the edges}

An alternative strategy to restore the gauge invariance
on the space manifold (\ref{eq: def space only manifold})
and the space-time manifold (\ref{eq: def space time manifold})
is to add to the action (\ref{1.1}) an action that cancels the
anomalous term (\ref{1.3}) acquired under the gauge transformation
(\ref{1.2}). In other words, the action
\begin{subequations}
\begin{equation}
S\:=
S^{\,}_{\mathrm{CS}}
+
S^{\,}_{\mathrm{edge}},
\end{equation}
where
\begin{equation}
\begin{split}
S^{\,}_{\mathrm{edge}}\:=&\,
\frac{1}{4\pi}\,
\int\mathrm{d}x\,\mathrm{d}t\,
\varphi^{(+)}\,
\left(
\partial^{\,}_{t}A^{(-)}_{1}
-
\partial^{\,}_{x}A^{(-)}_{0}
\right)
\\
&\,
+\frac{1}{4\pi}\,
\int\mathrm{d}x\,\mathrm{d}t\,
\varphi^{(-)}\,
\left(
\partial^{\,}_{t}A^{(+)}_{1}
-
\partial^{\,}_{x}A^{(+)}_{0}
\right),
\end{split}
\label{1.4a}
\end{equation}
is invariant under the gauge transformations defined by
Eqs.~(\ref{1.2}) and
\begin{equation}
\varphi^{(+)}\rightarrow
\varphi^{(+)}
+
\Lambda^{(+)},
\qquad
\varphi^{(-)}\rightarrow
\varphi^{(-)}+\Lambda^{(-)}.
\label{1.4b}
\end{equation}
\end{subequations}
The violation of the gauge symmetry in the bulk is exactly compensated
by the violation of the gauge symmetry at the edge. This is the
celebrated bulk-edge correspondence.\
\cite{Frohlich1,Frohlich2,Wen}

We now proceed to identifying the physical degrees of freedom at
the edge, by eliminating redundant degrees of freedom using
the symmetries of the edge action.

The path integral that defines the quantized theory along the edge
is to be performed over the 6 fields
\begin{equation}
\left\{
\varphi^{(-)},\varphi^{(+)},A^{(-)}_{0},A^{(+)}_{0},A^{(-)}_{1},A^{(+)}_{1}
\right\}.
\end{equation}
The symmetries of the action on the edge
follow from the gauge transformations
\begin{equation}
A^{(-)}_{0}\rightarrow
A^{(-)}_{0}
+
\partial^{\,}_{t}\chi^{(-)},
\qquad
A^{(-)}_{1}\rightarrow
A^{(-)}_{1}
+
\partial^{\,}_{x}\chi^{(-)}
\label{1.4c}
\end{equation}
and
\begin{equation}
A^{(+)}_{0}\rightarrow
A^{(+)}_{0}
+
\partial^{\,}_{t}\chi^{(+)},
\qquad
A^{(+)}_{1}\rightarrow
A^{(+)}_{1}
+
\partial^{\,}_{x}\chi^{(+)}.
\label{1.4d}
\end{equation}
The symmetry under
these transformations allows
to fix 2 degrees of freedom, say by demanding that
\begin{equation}
A^{(-)}_{0}=
v\,
A^{(+)}_{1},
\qquad
A^{(+)}_{0}=
v\,A^{(-)}_{1},
\label{1.4e}
\end{equation}
where the proportionality constant $v$ is an arbitrary real-valued
number carrying the dimension of velocity.
Insertion of the gauge-fixing conditions (\ref{1.4e}) into the edge
action (\ref{1.4a}) gives
\begin{equation}
\begin{split}
S^{\,}_{\mathrm{edge}}=&\,
\frac{1}{4\pi}\,
\int\mathrm{d}x\,\mathrm{d}t\,
\varphi^{(+)}\,
\left(
\partial^{\,}_{t} A^{(-)}_{1}
-
v\,
\partial^{\,}_{x}A^{(+)}_{1}
\right)
\\
&\,
+
\frac{1}{4\pi}\,
\int\mathrm{d}x\,\mathrm{d}t\,
\varphi^{(-)}\,
\left(
\partial^{\,}_{t}A^{(+)}_{1}
-
v\,
\partial^{\,}_{x}A^{(-)}_{1}
\right).
\end{split}
\label{1.4f}
\end{equation}

The path integral that defines the quantized theory along the edge
is now to be performed over the 4 fields
\begin{equation}
\left\{
\varphi^{(-)},\varphi^{(+)},A^{(-)}_{1},A^{(+)}_{1}
\right\}.
\end{equation}
The action on the edge is symmetric under the
residual gauge symmetry defined by
\begin{equation}
A^{(-)}_{1}\rightarrow
A^{(-)}_{1}
+
v\,
\partial^{\,}_{x}\zeta,
\qquad
A^{(+)}_{1}\rightarrow
A^{(+)}_{1}
+
\partial^{\,}_{t}\zeta
\label{1.4g}
\end{equation}
and
\begin{equation}
\varphi^{(+)}\rightarrow
\varphi^{(+)}
+
v\,
\partial^{\,}_{x}\xi,
\qquad
\varphi^{(-)}\rightarrow
\varphi^{(-)}
+
\partial^{\,}_{t}\xi,
\label{1.4h}
\end{equation}
provided $\zeta$ and $\xi$ satisfy the Klein-Gordon equations
\begin{equation}
(\partial^{2}_{t}-v^{2}\,\partial^{2}_{x})\,\zeta=0,
\qquad
(\partial^{2}_{t}-v^{2}\,\partial^{2}_{x})\,\xi=0,
\end{equation}
respectively. The functions $\zeta$ and $\xi$ that parametrize the residual
gauge symmetry obey the Klein-Gordon equation and as such can be decomposed
into a linear superposition of ingoing and outgoing waves,
\begin{subequations}
\begin{align}
\zeta(x,t)=
\zeta^{(+)}(x+v\,t)
+
\zeta^{(-)}(x-v\,t),
\label{1.4i}
\\
\xi(x,t)=
\xi^{(+)}(x+v\,t)
+
\xi^{(-)}(x-v\,t).
\label{1.4j}
\end{align}
\end{subequations}
The components
$\zeta^{(+)},\xi^{(+)}$
and
$\zeta^{(-)},\xi^{(-)}$
when $v>0$ are also known as left- and right- moving waves
or as chiral and anti-chiral waves,
respectively.

The functions $\zeta$ and $\xi$ are not the only ones obeying the
Klein-Gordon equation. So do the dynamical fields
$A^{(\pm)}_{1}$
and
$\varphi^{(\pm)}$,
\begin{equation}
(\partial^{2}_{t}-v^{2}\,\partial^{2}_{x})\,
A^{(\pm)}_{1}=0,
\qquad
(\partial^{2}_{t}-v^{2}\,\partial^{2}_{x})\,
\varphi^{(\pm)}=0,
\end{equation}
as follows from the equations of motion
derived from the action on the edge (\ref{1.4f}).
Correspondingly, the fields
$A^{(\pm)}_{1}$
and
$\varphi^{(\pm)}$
also obey an additive decomposition into
chiral and anti-chiral components. This observation allows to
impose the gauge-fixing condition
\begin{equation}
A^{(-)}_{1}=
-
\partial^{\,}_{x}\,\varphi^{(-)},
\qquad
A^{(+)}_{1}=
-\partial^{\,}_{x}\,\varphi^{(+)}.
\label{eq: final gauge fixing on edge}
\end{equation}

Implementing the condition
(\ref{eq: final gauge fixing on edge})
in the action on the edge (\ref{1.4f})
delivers
\begin{widetext}
\begin{equation}
S^{\,}_{\mathrm{edge}}=
\frac{1}{4\pi}\,
\int\mathrm{d}x\,\mathrm{d}t\,
\left(
\partial^{\,}_{t}\varphi^{(-)}\partial^{\,}_{x}\varphi^{(+)}
+
\partial^{\,}_{t}\varphi^{(+)}\partial^{\,}_{x}\varphi^{(-)}
-
v\,
\partial^{\,}_{x}\varphi^{(+)}\partial^{\,}_{x}\varphi^{(+)}
-
v\,
\partial^{\,}_{x}\varphi^{(-)}\partial^{\,}_{x}\varphi^{(-)}
\right)
\label{1.9 bis}
\end{equation}
\end{widetext}
in agrement with Eq.~(\ref{1.9}). The derivation
of Eq.~(\ref{1.9 bis}) is the one that we will extend to the case
when  Rashba spin-orbit coupling is present.

\subsection{Including Rashba Terms - BRST Approach}

The effective action in the presence of Rashba spin-orbit coupling is
\begin{widetext}
\begin{equation}
I[A^{(+)},A^{(-)}]=
-\frac{1}{4\pi}\,
\int\mathrm{d}^{3}x\,
\left[
\epsilon^{\mu\nu\lambda}\,
\left(
A^{(+)}_{\mu}\,\partial^{\,}_{\nu}A^{(-)}_{\lambda}
+
A^{(-)}_{\mu}\,\partial^{\,}_{\nu}A^{(+)}_{\lambda}
\right)
-
\frac{\lambda^{2}_{\mathrm{R}}}{|\eta|}\,
(A^{(-)}_{0})^{2}
\right].
\label{4.1}
\end{equation}
\end{widetext}
Owing to the term $(A^{(-)}_{0})^{2}$, we no longer have the full
gauge symmetry (\ref{1.2})
(nor the Lorentz symmetry),
that was used to
establish the bulk-edge correspondence in the case of the
doubled Chern-Simons theory.
On the other hand, we are in the situation where
$|\lambda^{\,}_{\mathrm{R}}|\ll|\eta|$.
Hence, the existence of the gap in the bulk is not affected
by switching on adiabatically the Rashba spin-orbit coupling.
The topological stability of the parity in the number of pairs of
gapless helical edge states implies that at least one pair
remains gapless when $|\lambda^{\,}_{\mathrm{R}}|\ll|\eta|$.
Our task is now to understand
how to get the edge states from the effective field theory
(\ref{4.1}).  It is natural to expect that some weaker symmetry
replaces the gauge symmetry of
Sec.\ \ref{subsec: Pure Chern-Simons Theory}.

We start with a manifold without boundaries.  The central point
of our construction is that the Rashba term can be interpreted
as a gauge-fixing term for the field $A^{(-)}_{\mu}$. If so, the action
(\ref{4.1}) is to be thought of as a doubled Chern-Simons action augmented
by a gauge-fixing term,  which we may implement
through the Faddeev-Popov procedure\cite{Faddeev-Popov}, as we now show.
To this end, we define the ghost action
\begin{subequations}
\begin{equation}
S^{\,}_{\mathrm{ghost}}\:=
-\frac{1}{4\pi}\,
\int\mathrm{d}^{3}x\,
\bar{C}\,\partial^{\,}_{t}C,
\label{4.2}
\end{equation}
where $\bar{C}$ and $C$ enter as
Grassmann-valued ghosts fields in the partition function.
The augmented action is
\begin{equation}
S\:=
I
+
S^{\,}_{\mathrm{ghost}}.
\label{eq: def augmented action}
\end{equation}
We may then write
\begin{align}
Z\:=&\,
\int\mathcal{D}A\;
e^{+\mathrm{i}I}
\nonumber\\
\propto&\,
\int\mathcal{D}A\,
\int\mathcal{D}C\,\mathcal{D}\bar{C}\;
e^{\mathrm{i}(I+S^{\,}_{\mathrm{ghost}})},
\label{4.5}
\end{align}
\end{subequations}
for there is no coupling between the gauge and ghost fields
and the integration over the ghosts just produces
a constant multiplicative factor
that can be absorbed in the integration measure,
\begin{equation}
\int\mathcal{D}C\,\mathcal{D}\bar{C}\;
e^{\mathrm{i}S^{\,}_{\mathrm{ghost}}}=
\mathrm{constant}.
\end{equation}

The partition function (\ref{4.5})
is independently invariant (as the action changes by a a total derivative)
under the gauge transformation
\begin{equation}
A^{(+)}_{\mu}\rightarrow
A^{(+)}_{\mu}
+
\partial^{\,}_{\mu}\Lambda^{(+)}
\label{4.3}
\end{equation}
for the gauge field $A^{(+)}_{\mu}$
and the BRST transformations\cite{BRS,T,Weinberg,Polchinski}
\begin{equation}
\begin{split}
&
A^{(-)}_{\mu}\rightarrow
A^{(-)}_{\mu}
+
\theta\,
\partial^{\,}_{\mu}C,
\\
&
\bar{C}\rightarrow
\bar{C}
+
2\,
\frac{\lambda^{2}_{\mathrm{R}}}{|\eta|}\,
\theta\,
A^{(-)}_{0},
\\
&
C\rightarrow
C,
\end{split}
\label{4.4}
\end{equation}
for the gauge field $A^{(-)}_{\mu}$
and for the pair $\bar{C}$ and $C$ of ghost fields.
Here, $\theta$ is a global Grassmannian parameter of the BRST transformation.
Observe that the transformation of
the $A^{(-)}_{\mu}$ is essentially a gauge transformation with parameter
$\theta\,C$. The form of the BRST transformation shows that when
the Rashba coupling constant $\lambda^{\,}_{\mathrm{R}}\rightarrow 0$,
the ghosts fields no longer transform and
we recover the transformations (\ref{1.2})
with the identification $\Lambda^{(-)}\equiv\theta\,C$.
That is the reason for which we do not need to invoke the ghosts fields
in the doubled Chern-Simons theory.
The BRST approach in the up-down basis is discussed in the appendix
\ref{appendixC}.

On the one hand, the inclusion of the ghost action in
(\ref{4.1}) in our effective field theory is innocuous, for the
ghost fields can be thought of as being hidden, i.e., integrated
out, and it is a mere matter of convenience to make them explicit.
On the other hand, the inclusion of the ghost action is important to
understand how the bulk effective theory (\ref{4.1}) delivers gapless
edge states.

In the presence of the space manifold (\ref{eq: def space only manifold})
and the space-time manifold (\ref{eq: def space time manifold}),
the action (\ref{eq: def augmented action})
is no longer invariant under the gauge and BRST
transformations (\ref{4.3}) and (\ref{4.4}), respectively.
The action (\ref{eq: def augmented action}
changes under the transformations (\ref{4.3}) and (\ref{4.4})
by the boundary action
\begin{equation}
\begin{split}
\delta S=&\,
-\frac{1}{4\pi}\,
\int\mathrm{d}x\,\mathrm{d}t\,
\Lambda^{(+)}\,
\left(
\partial^{\,}_{t}A^{(-)}_{1}
-
\partial^{\,}_{x}A^{(-)}_{0}
\right)
\Big{|}^{\,}_{y=0}
\\
&\,
-\frac{1}{4\pi}\,
\int\mathrm{d}x\,\mathrm{d}t\,
\theta\,
C\,
\left(
\partial^{\,}_{t}A^{(+)}_{1}
-
\partial^{\,}_{x}A^{(+)}_{0}
\right)
\Big{|}^{\,}_{y=0}.
\label{4.6}
\end{split}
\end{equation}

Invariance under the transformations
(\ref{4.3}) and (\ref{4.4})
is achieved by the partition function with the action
\begin{subequations}
\begin{equation}
S^{\,}_{\mathrm{tot}}\:=
I
+
S^{\,}_{\mathrm{ghost}}
+
S^{\,}_{\mathrm{edge}},
\end{equation}
where
\begin{equation}
\begin{split}
S^{\,}_{\mathrm{edge}}\:=&\,
\frac{1}{4\pi}\,
\int\mathrm{d}x\,\mathrm{d}t\,
\varphi^{(+)}\,
\left(
\partial^{\,}_{t}A^{(-)}_{1}
-
\partial^{\,}_{x}A^{(-)}_{0}
\right)
\\
&\,
+
\frac{1}{4\pi}\,
\int\mathrm{d}x\,\mathrm{d}t\,
\varphi^{(-)}\,
\left(
\partial^{\,}_{t}A^{(+)}_{1}
-
\partial^{\,}_{x}A^{(+)}_{0}
\right),
\end{split}
\label{4.7}
\end{equation}
and the edge fields $\varphi^{(\pm)}$ transform according to the law
\begin{equation}
\varphi^{(+)}\rightarrow
\varphi^{(+)}
+
\Lambda^{(+)},
\qquad
\varphi^{(-)}\rightarrow
\varphi^{(-)}
+
\theta\,
C.
\label{4.8}
\end{equation}
\end{subequations}

The action on the edge (\ref{4.7}) is none but
the action (\ref{1.4a}). The gauge fixing
from Sec.\ \ref{subsec: Restoring gauge symmetry by adding ... the edges}
is applicable and delivers
\vskip 60 true pt
\begin{widetext}
\begin{equation}
S^{\,}_{\mathrm{edge}}=
\frac{1}{4\pi}\,
\int\mathrm{d}x\,\mathrm{d}t\,
\left(
\partial^{\,}_{t}\varphi^{(-)}\,
\partial^{\,}_{x}\varphi^{(+)}
+
\partial^{\,}_{t}\varphi^{(+)}\,
\partial^{\,}_{x}\varphi^{(-)}
-
v\,
\partial^{\,}_{x}\varphi^{(+)}\,
\partial^{\,}_{x}\varphi^{(+)}
-
v\,
\partial^{\,}_{x}\,\varphi^{(-)}\,
\partial^{\,}_{x}\varphi^{(-)}
\right),
\label{4.9}
\end{equation}
\end{widetext}
in agrement with Eqs.~(\ref{1.9}) and (\ref{1.9 bis}).
Hence, we have shown that the existence of a single pair of
gapless helical edge states in the quantum-spin Hall effect is
robust to the adiabatic switching of a Rashba spin-orbit coupling
$|\lambda^{\,}_{\mathrm{R}}|\ll|\eta|$.

\section{Discussion}
\label{sec: Discussions}

In this work, we have obtained the low-energy and long-wave length
effective field theory that encodes the Kane-Mele model
with a dominant intrinsic spin-orbit coupling and a subdominant
Rashba spin-orbit coupling at vanishing uniform and staggered
chemical potentials \cite{footnote: Kane-Mele-Hubbard}.
Without Rashba spin-orbit coupling the
fermionic Lagrangian density (\ref{1}) has the Lorentz, $U(1)\times U(1)$
gauge, and time-reversal symmetries. All these symmetries are encoded by the
doubled Chern-Simons effective
Lagrangian density (\ref{eq: def double CS Lagrangian})
that follows from integrating out the fermions to lowest order in
a gradient expansion.

The effect of the Rashba coupling is the additive correction
$(A^{(-)}_{0})^{2}$ to the double
Chern-Simons Lagrangian that breaks the
gauge invariance of $A_{\mu}^{(-)}$ as well as the Lorentz symmetry of
the theory, while preserving time-reversal symmetry.
On the other hand, the
gauge invariance of $A^{(+)}_{\mu}$ is preserved due to the
conservation of electric charge.

The requirement of gauge invariance when the physics in the bulk
and at the boundaries are treated on equal footing
is the ingredient sufficient to establish the
bulk-edge correspondence for the doubled Chern-Simons with the
Lagrangian density (\ref{eq: def double CS Lagrangian}).
However, the correction due to the Rashba coupling partially breaks
the $U(1)\times U(1)$ gauge invariance down to $U(1)$.
Nevertheless, topological arguments constructed from the Bloch states
associated with the band electrons guarantee the existence of
an odd number of pairs of gapless helical edge states whenever
$|\lambda^{\,}_{\mathrm{R}}|$ is small compared to the spin-orbit
coupling $|\eta|$. Thus, the question is how to determine the bulk-edge
correspondence, in this case without the $U(1)\times U(1)$ gauge
invariance.  Our strategy was to interpret the correction
$\sim(A^{(-)}_{0})^{2}$ as a gauge fixing term for the $A^{(-)}_{\mu}$
field. In this way, in replacement of the
$U(1)$ gauge (residual spin) symmetry,
we find a BRST symmetry
after the appropriate ghost action is accounted for.  For a manifold with a
boundary, the   BRST symmetry
delivers the bulk-edge correspondence
leading to a pair of gapless helical edge states.
As there is no interaction between ghost and
gauge fields and as the BRST transformation reduces
to the usual gauge transformations in the limit
$\lambda^{\,}_{\mathrm{R}}\rightarrow 0$,
the $U(1)\times U(1)$ gauge symmetry of the doubled Chern-Simons action
is recovered in the $\lambda^{\,}_{\mathrm{R}}\rightarrow 0$ limit.

Having succeeded in establishing the bulk-edge correspondence in
the presence of the Rashba spin-orbit coupling using the BRST symmetry,
we now turn the discussion to open problems. The approach we proposed needs
to be extended to more general situations in which time-reversal
symmetry is preserved.  This is the case when we consider arbitrary
vectors $V^{\,}_{i}$ and $W^{\,}_{i}$ ($W^{\,}_{i}=V^{\ast}_{i}$ and
$V^{\,}_{0}=W^{\,}_{0}=0$).
From our one-loop calculations, we infer that
the low energy effective Lagrangian density is
\begin{equation}
\begin{split}
\mathcal{L}=&\,
\epsilon^{\mu\nu\lambda}\,
\left(
A^{(+)}_{\mu}\,\partial^{\,}_{\nu}A^{(-)}_{\lambda}
+
A^{(-)}_{\mu}\,\partial^{\,}_{\nu}A^{(+)}_{\lambda}
\right)
\\
&\,
-
\frac{\lambda^{2}_{\mathrm{R}}}{|\eta|}\,
f^{\mu\nu}\,
A^{(-)}_{\mu}\,
A^{(-)}_{\nu},
\end{split}
\label{7.6}
\end{equation}
where $f^{\mu\nu}$ is an arbitrary real-valued symmetric matrix
with $f^{0i}=0$ that can be read from Eq.\ (\ref{34}), i.e.,
\begin{equation}
f^{\mu\nu}\propto
\left[
\frac12\, g^{\mu\nu}\,(V\cdot W)
-
\frac12\,
(
V^{\mu}\,W^{\nu}
+
V^{\nu}W^{\mu}
)
\right].
\label{7.23}
\end{equation}
One verifies that the BRST procedure cannot be directly applied
to this more general situation.  This is so because the correction
$f^{\mu\nu}\,A^{(-)}_{\mu}\,A^{(-)}_{\nu}$ does not correspond to a gauge
fixing term. In the sense of the gauge fixing, it fixes more
components than allowed by gauge invariance.  Thus, a remaining
problem is how to determine the bulk-edge correspondence in this
situation.

It is encouraging to view the problem from the following
perspective.  We do know that the bulk, described by
Eqs.~(\ref{7.6}) and~(\ref{7.23}), does have gapless edge modes,
because it realizes a $\mathbb{Z}^{\,}_{2}$ topological insulator.
We succeeded in uncovering a weaker symmetry than the $U(1)$
gauge symmetry to establish the bulk-edge correspondence
when a small Rashba spin-orbit coupling is present.
The BRST symmetry is perhaps sufficient, yet not necessary, to establish
the bulk-edge correspondence, in which case a weaker condition than BRST
would be the guarantor for the bulk-edge correspondence.

\medskip
\section{Acknowledgments}

P.R.S.G. thanks Marcelo Gomes for useful discussions and
Funda\c{c}\~ao de Amparo a Pesquisa do Estado de S\~ao Paulo (FAPESP)
for the financial support. C.M. thanks the Condensed
Matter Theory Visitors Program at Boston University
for support.
This work was supported by DOE Grant
DEF-06ER46316 (P-.H.H. and C.C.).

\appendix

\section{Some useful properties of Dirac matrices}\label{apendiceA}

In this appendix, we recall some properties of Dirac matrices useful
in the calculation of the Feynman diagrams contributing to the effective
action. The first property is the product of two Dirac matrices, that
can be decomposed as
\begin{align}
\Gamma^{\mu}\Gamma^{\nu}=&\,
\frac12\,
\{\Gamma^{\mu},\Gamma^{\nu}\}
+
\frac12\,
[\Gamma^{\mu},\Gamma^{\nu}]
\nonumber\\
=&\,
g^{\mu\nu}-
\mathrm{i}\epsilon^{\mu\nu\rho}\,
\Gamma^{\,}_{\rho},
\label{a1}
\end{align}
where we used
$[\Gamma^{\mu},\Gamma^{\nu}]=-2\mathrm{i}\epsilon^{\mu\nu\rho}\,\Gamma^{\,}_{\rho}$,
with the convention $\epsilon^{012}\equiv1$.  This property enable us
to reduce the number of Dirac matrices in products with several
matrices.  From Eq.\ (\ref{a1}),
we may easily obtain the trace of products
of Dirac matrices
\begin{equation}
\text{Tr}\,(\Gamma^{\mu}\,\Gamma^{\nu})=4g^{\mu\nu},
\label{a2}
\end{equation}
\begin{equation}
\text{Tr}\,
\left(\Gamma^{\mu}\,\Gamma^{\nu}\,\Gamma^{\rho}\right]=
-4\mathrm{i}\epsilon^{\mu\nu\rho},
\label{a3}
\end{equation}
and
\begin{equation}
\text{Tr}\,
\left[\Gamma^{\mu}\,\Gamma^{\nu}\,\Gamma^{\rho}\,\Gamma^{\sigma}\right]=
4(g^{\mu\nu}\,g^{\rho\sigma}-g^{\mu\rho}\,g^{\nu\sigma}+g^{\mu\sigma}\,g^{\rho\nu}).
\label{a4}
\end{equation}
A helpful property involving the Levi-Civita tensor is
\begin{equation}
\epsilon^{\mu\nu\sigma}\,\epsilon^{\alpha\beta}\,_{\sigma}=
g^{\mu\alpha}\,g^{\nu\beta}-g^{\mu\beta}\,g^{\nu\alpha}.
\label{a5}
\end{equation}

\section{Calculation of Diagrams}
\label{apendiceB}

This appendix is dedicated to the calculation of Feynman diagrams
involved in the determination of the low energy effective field theory
underlying the Kane-Mele Hamiltonian.

We will discuss a procedure to obtain the expression for the tensor
$A^{\mu\nu\alpha\beta}$ given in Eq.\ (\ref{22}).
At first, we need to deal with the following trace of Dirac matrices
\begin{equation}
\text{Tr}\,
\left[
\Gamma^{\mu}\,
(\slash\!\!\!{k}+\eta)\,
\Gamma^{\alpha}\,
(\slash\!\!\!{k}-\eta)\,
\Gamma^{\beta}\,
(\slash\!\!\!{k}+\eta)
\Gamma^{\nu}\,
(\slash\!\!\!{k}+\eta)
\right].
\label{15}
\end{equation}
Note that we can reduce the number of Dirac matrices in this product
by using the algebra of Dirac matrices (\ref{s1.11})
and the commutator
$[\Gamma^{\mu},\Gamma^{\nu}]=-2\mathrm{i}\epsilon^{\mu\nu\rho}\,\Gamma^{\,}_{\rho}$.
So we have
\begin{equation}
(\slash\!\!\!{k}+\eta)\,
\Gamma^{\mu}\,
(\slash\!\!\!{k}+\eta)=
2k^{\mu}\,
(\slash\!\!\!{k}+\eta)
-
(k^{2}-\eta^{2})\,
\Gamma^{\mu}
\label{16}
\end{equation}
and
\begin{equation}
(\slash\!\!\!{k}+\eta)\,
\Gamma^{\mu}\,
(\slash\!\!\!{k}-\eta)=
(k^{2}-\eta^{2})\,
\Gamma^{\mu}
+
2\mathrm{i}(\slash\!\!\!{k}+\eta)
\epsilon^{\rho\mu\sigma}\,
k^{\,}_{\rho}\,
\Gamma_{\sigma}.
\label{17}
\end{equation}
With this, the trace in (\ref{15}) becomes
\begin{widetext}
\begin{align}
\text{Tr}\,
\Big[&\!\!
2(k^{2}-\eta^{2})\,
k^{\nu}\,
k^{\,}_{\rho}\,
\Gamma^{\mu}\,
\Gamma^{\alpha}\,
\Gamma^{\beta}\,
\Gamma^{\rho}
-
(k^{2}-\eta^{2})^{2}\,
\Gamma^{\mu}\,
\Gamma^{\alpha}\,
\Gamma^{\beta}\,
\Gamma^{\nu}
+
4\mathrm{i}
k^{\,}_{\lambda}\,
k^{\,}_{\delta}\,
k^{\nu}\,
k^{\,}_{\rho}\,
\epsilon^{\rho\alpha}\,_{\sigma}
\Gamma^{\mu}\,
\Gamma^{\lambda}\,
\Gamma^{\sigma}\,
\Gamma^{\beta}\,
\Gamma^{\delta}
\nonumber\\
&\,+
4\mathrm{i}\eta^{2}\,
k^{\nu}\,
k^{\,}_{\rho}\,
\epsilon^{\rho\alpha}\,_{\sigma}\,
\Gamma^{\mu}\,
\Gamma^{\sigma}\,
\Gamma^{\beta}
-
2\mathrm{i}
(k^{2}-\eta^{2})\,
k^{\,}_{\lambda}\,
k^{\,}_{\rho}\,
\epsilon^{\rho\alpha}\,_{\sigma}\,
\Gamma^{\mu}\,
\Gamma^{\lambda}\,
\Gamma^{\sigma}\,
\Gamma^{\beta}\,
\Gamma^{\nu}
\Big],
\label{18}
\end{align}
where we discarded terms with an odd number of loop momentum that
vanish when integrated. For the terms involving a product of five
Dirac matrices, it is convenient to use the decomposition
$\Gamma^{\sigma}\,\Gamma^{\beta}=
g^{\sigma\beta}-\mathrm{i}\epsilon^{\sigma\beta\eta}\,\Gamma^{\,}_{\eta}$
in order to reduce the number of matrices.  After that, eliminating
the terms involving two Levi-Civita with one contracted index by means
of the relation (\ref{a5}), we obtain the following result for the
trace
\begin{align}
&
8(k^{2}-\eta^{2})\,
(
g^{\mu\alpha}\,k^{\nu}\,k^{\beta}
-
g^{\mu\beta}\,k^{\nu}\,k^{\alpha}
+
g^{\alpha\beta}\,k^{\mu}\,k^{\nu})
-
4(k^{2}-\eta^{2})^{2}\,
(g^{\mu\alpha}\,g^{\beta\mu}-g^{\mu\beta}\,g^{\alpha\nu}+g^{\mu\nu}\,g^{\alpha\beta})
\nonumber\\
&
+
32\,
k^{\mu}\,
k^{\nu}\,
k^{\alpha}\,
k^{\beta}
-
16\,
k^{2}\,
(
g^{\mu\alpha}\,k^{\beta}\,k^{\nu}
+
g^{\alpha\beta}\,k^{\mu}\,k^{\nu}
)
-
16\,\eta^{2}
(
g^{\alpha\beta}\,k^{\mu}\,k^{\nu}
-
g^{\mu\alpha}\,k^{\beta}\,k^{\nu}
)
+
8\,
(k^{2}-\eta^{2})\,
\epsilon^{\rho\alpha\beta}\,
\epsilon^{\sigma\mu\nu}\,
k^{\,}_{\rho}\,
k^{\,}_{\sigma}
\nonumber\\
&
-
8\,
(
k^{2}-\eta^{2}
)\,
(
g^{\nu\alpha}\,k^{\mu}\,k^{\beta}
-
g^{\mu\alpha}\,k^{\nu}\,k^{\beta}
+
g^{\mu\nu}\,k^{\alpha}\,k^{\beta}
)
+
8\,k^{2}\,
(k^{2}-\eta^{2})\,
g^{\alpha\beta}\,
g^{\mu\nu}.
\label{19}
\end{align}
We can take advantage of the Lorentz invariance to do the following replacements
\begin{equation}
k^{\mu}\,
k^{\nu}\rightarrow
\frac{1}{D}\,
g^{\mu\nu}\,
k^{2}
\label{20}
\end{equation}
and
\begin{equation}
k^{\mu}\,
k^{\nu}\,
k^{\alpha}\,
k^{\beta}\rightarrow
\frac{1}{D\,(D+2)}\,
(
g^{\mu\nu}\,g^{\alpha\beta}
+
g^{\mu\alpha}\,g^{\nu\beta}
+
g^{\mu\beta}\,g^{\nu\alpha}
)\,
(k^{2})^{2},
\label{21}
\end{equation}
that are valid under the momentum integration. For our case $D=3$.
The result is
\begin{align}
A^{\mu\nu\alpha\beta}=&\,
\frac83\,
(
g^{\mu\alpha}\,g^{\nu\beta}
-
g^{\mu\beta}\,g^{\nu\alpha}
+
g^{\alpha\beta}\,g^{\mu\nu}
)\,
J^{1}_{3}
-
4\,
(
g^{\mu\alpha}\,
g^{\beta\mu}
-
g^{\mu\beta}\,
g^{\alpha\nu}
+
g^{\mu\nu}\,
g^{\alpha\beta}
)\,
J^{0}_{2}
\nonumber\\
&\,
+
\frac{32}{15}\,
(
g^{\mu\nu}\,g^{\alpha\beta}
+
g^{\mu\alpha}\,g^{\nu\beta}
+
g^{\mu\beta}\,g^{\nu\alpha}
)\,
J^{2}_{4}
-
\frac{16}{3}\,
(
g^{\mu\alpha}\,g^{\beta\nu}
+
g^{\alpha\beta}\,g^{\mu\nu})\,
J^{2}_{4}
\nonumber\\
&\,
-
16\,\eta^{2}\,
(
g^{\alpha\beta}\,g^{\mu\nu}
-
g^{\mu\alpha}\,g^{\beta\nu}
)\,
J^{1}_{4}
+
\frac83\,
(
g^{\alpha\mu}\,g^{\beta\nu}
-
g^{\alpha\nu}\,g^{\beta\mu}
)
J^{1}_{3}
\nonumber\\
&\,
-
\frac83\,
(
g^{\nu\alpha}\,g^{\mu\beta}
-
g^{\mu\alpha}\,g^{\nu\beta}
+
g^{\mu\nu}\,g^{\alpha\beta}
)\,
J^{1}_{3}
+
8\,
g^{\alpha\beta}\,g^{\mu\nu}\,J^{1}_{3},
\label{b1}
\end{align}
where we defined the integral $J^{P^{\,}_{2}}_{P^{\,}_{1}}$ to be
\begin{align}
J_{P_1}^{P_2}\equiv&\,
\int\frac{\mathrm{d}^{D}k}{(2\pi)^{3}}\,
\frac{(k^{2})^{P^{\,}_{1}}}{(k^{2}-\eta^{2}+\mathrm{i}\epsilon)^{P^{\,}_{2}}}
\nonumber\\
=&\,
\mathrm{i}(-1)^{P^{\,}_{1}-P^{\,}_{2}}\,
\frac{\Omega^{\,}_{D}}{(2\pi)^{d}}\,
\frac{1}{\Gamma(P^{\,}_{2})}\,
\Gamma\left(\frac{2P^{\,}_{1}+D}{2}\right)\,
\Gamma\left(\frac{2P^{\,}_{2}-2P^{\,}_{1}-D}{2}\right)\,
\frac{1}{|\eta|^{2P^{\,}_2{}-2P^{\,}_{1}-D}},
\label{b2}
\end{align}
with $\Omega_D\equiv \frac{2\pi^{D/2}}{\Gamma(D/2)}$.
Using this result in (\ref{b1}), we obtain (\ref{22}),
\begin{equation}
A^{\mu\nu\alpha\beta}=
\frac{\mathrm{i}}{12\,\pi\,|\eta|}\,
\left(
g^{\mu\nu}\,g^{\alpha\beta}
-
g^{\mu\alpha}\,g^{\nu\beta}
-
g^{\mu\beta}\,g^{\nu\alpha}
\right).
\label{b3}
\end{equation}
A second type of 4-index tensor useful to deal with the diagrams in
Figs.\ \ref{f7} and \ref{f8} is
\begin{equation}
B^{\mu\nu\alpha\beta}\equiv
\int\frac{\mathrm{d}^{3}k}{(2\pi)^{3}}\,
\text{Tr}\,
\left(
\Gamma^{\mu}\,
S^{u}(k)\,
\Gamma^{\alpha}\,
S^{d}(k)\,
\Gamma^{\nu}\,
S^{d}(k)\,
\Gamma^{\beta}\,
S^{u}(k)
\right).
\label{b4}
\end{equation}
\end{widetext}
The trace we need to consider is
\begin{equation}
\text{Tr}\,
\left[
(\slash\!\!\!{k}+\eta)\,
\Gamma^{\mu}\,
(\slash\!\!\!{k}+\eta)\,
\Gamma^{\alpha}\,
(\slash\!\!\!{k}-\eta)\,
\Gamma^{\nu}\,
(\slash\!\!\!{k}-\eta)\,
\Gamma^{\beta}
\right].
\label{b5}
\end{equation}
Observe that this expression is more symmetric than Eq.\ (\ref{15}). In
this case, we can use
\begin{equation}
(\slash\!\!\!{k}+\eta)\,
\Gamma^{\mu}\,
(\slash\!\!\!{k}+\eta)=
2\,
k^{\mu}\,
(\slash\!\!\!{k}+\eta)
-
(k^{2}-\eta^{2})\,
\Gamma^{\mu}
\label{b6}
\end{equation}
and
\begin{equation}
(\slash\!\!\!{k}-\eta)\,
\Gamma^{\nu}(\slash\!\!\!{k}-\eta)=
2\,
k^{\nu}\,
(\slash\!\!\!{k}-\eta)
-
(k^{2}-\eta^{2})\,
\Gamma^{\nu}.
\label{b7}
\end{equation}
After that, by following essentially the same steps that yielded
Eq.\ (\ref{b2}), we obtain
\begin{equation}
B^{\mu\nu\alpha\beta}=
-\frac{\mathrm{i}}{6\pi\,|\eta|}\,
\left(
g^{\mu\nu}\,g^{\alpha\beta}
-
g^{\mu\alpha}\,g^{\nu\beta}
-
g^{\mu\beta}\,g^{\nu\alpha}
\right).
\label{b8}
\end{equation}


\section{BRST approach in the up-down basis}\label{appendixC}

We will discuss the BRST approach with the gauge fields in the up-down basis. The action (\ref{4.1}) written in terms of
$A_{\mu}^{u}$ and $A_{\mu}^{d}$ fields is
\begin{widetext}
\begin{equation}
S^{\,}_{\mathrm{eff}}=
-\frac{1}{2\pi}\,
\int\mathrm{d}^{3}x
\left[
\epsilon^{\mu\nu\lambda}\,
\left(
A^{u}_{\mu}\,\partial^{\,}_{\nu}A^{u}_{\lambda}
-
A^{d}_{\mu}\,\partial^{\,}_{\nu}A^{d}_{\lambda}
\right)
-
\frac{\lambda^{2}_{\mathrm{R}}}{2|\eta|}\,
(A^{u}_{0}-A^{d}_{0})^{2}
\right].
\label{c1}
\end{equation}
\end{widetext}
The Rashba term breaks the gauge invariance of both $A_{\mu}^{u}$ and $A_{\mu}^{d}$ fields.
So it is natural to expect the existence of two types of gauge fields ($C^{u}$, $\bar{C}^{u}$) and ($C^{d}$, $\bar{C}^{d}$).
The ghost action is
\begin{equation}
S^{\,}_{\mathrm{ghost}}=
-\frac{1}{2\pi}\,
\int\mathrm{d}^{3}x\,
\left(
\bar{C}^{u}\,\partial^{\,}_{0}C^{u}
+
\bar{C}^{d}\partial^{\,}_{0}C^{d}
\right).
\label{c2}
\end{equation}
The BRST transformations are
\begin{equation}
\begin{split}
&
A^{u}_{\mu}\rightarrow
A^{u}_{\mu}
+
\theta\,
\partial^{\,}_{\mu}C^{u},
\\
&
C^{u}\rightarrow
C^{u},
\\
&
\bar{C^{u}}\rightarrow
\bar{C}^{u}
+
\frac{\lambda^{2}_{\mathrm{R}}}{|\eta|}\,
\theta\,
(A^{u}_{0}-A^{d}_{0}),
\end{split}
\label{c3}
\end{equation}
and
\begin{equation}
\begin{split}
&
A^{d}_{\mu}\rightarrow
A^{d}_{\mu}
+
\theta\,
\partial^{\,}_{\mu}C^{d},
\\
&
C^{d}\rightarrow
C^{d},
\\
&
\bar{C^{d}}\rightarrow
\bar{C}^{d}
-
\frac{\lambda^{2}_{\mathrm{R}}}{|\eta|}\,
\theta\,(A^{u}_{0}-A^{d}_{0}).
\end{split}
\label{c4}
\end{equation}
Under these transformations the variation of the action (\ref{c1}) is
the surface term
\begin{equation}
\delta S^{\,}_{\mathrm{eff}}=
-\frac{1}{2\pi}\,
\int\mathrm{d}^{3}x\,
\partial^{\,}_{\mu}
\left[
\epsilon^{\mu\nu\lambda}\,
\left(
\theta\,C^{u}\,\partial^{\,}_{\nu}A^{u}_{\lambda}
-
\theta\,C^{d}\,\partial^{\,}_{\nu}A^{d}_{\lambda}
\right)
\right].
\label{c5}
\end{equation}
If we choose a manifold with a boundary at $y=0$, as before,
we obtain the edge contribution
\begin{align}
\delta S^{\,}_{\mathrm{edge}}=&\,
-\frac{1}{2\pi}\int\mathrm{d}x\,\mathrm{d}t\,
\left[
\theta\,
C^{u}\,
(
\partial^{\,}_{t}A^{u}_{1}
-
\partial^{\,}_{x}\,A^{u}_{0}
)
\right.
\nonumber\\
&\,
-
\left.
\theta\,C^{d}\,
(
\partial^{\,}_{t}A^{d}_{1}
-
\partial^{\,}_{x}A^{d}_{0}
)
\right].
\label{c6}
\end{align}
By analyzing the symmetries of the edge we can find the edge states.

We can connect the above construction with the discussion in the text
by passing to the $\pm$ basis.  We introduce the gauge fields
$A^{(\pm)}_{\mu}\equiv A^{u}_{\mu}\pm A^{d}_{\mu}$ and similar
definitions for the ghost fields $C^{(\pm)}\equiv C^{u}\pm C^{d}$ and
$\bar{C}^{(\pm)}\equiv \bar{C}^{u}\pm \bar{C}^{d}$. With this, the gauge
action is given by Eq.\ (\ref{4.1}) whereas the ghost action (\ref{c2})
becomes
\begin{equation}
S^{\,}_{\mathrm{ghost}}\propto
\int\mathrm{d}^{3}x\,
\left(
\bar{C}^{(+)}\,\partial^{\,}_{0}C^{(+)}
+
\bar{C}^{(-)}\,\partial^{\,}_{0}C^{(-)}
\right).
\label{c7}
\end{equation}
This is not the action we constructed in Eq.\ (\ref{4.2}). We have the
presence of additional ghosts degrees of freedom.  However, according
to Eqs.\ (\ref{c3}) and (\ref{c4}), we see that the transformations of the
ghosts fields $C^{(\pm)}$ and $\bar{C}^{(\pm)}$ are
\begin{equation}
\delta \bar{C}^{(+)}=0,
\qquad
\delta C^{(+)}=0
\label{c8}
\end{equation}
and
\begin{equation}
\delta \bar{C}^{(-)}=
\frac{2\lambda^{2}_{\mathrm{R}}}{|\eta|}\,
\theta\, (A^{u}_{0}-A^{d}_{0}) ,
\qquad
\delta C^{(-)}=0,
\label{c9}
\end{equation}
besides the transformation of the gauge fields $\delta
A^{(\pm)}_{\mu}=\theta\,\partial^{\,}_{\mu}C^{(\pm)}$.  The ghosts
fields $\bar{C}^{(+)}$ and ${C}^{(+)}$ do not transform and hence the
contribution $\bar{C}^{(+)}\partial^{\,}_{0}C^{(+)}$ can be discarded
from the action (\ref{c7}), yielding the desired result with the
identifications $\bar{C}^{(-)}\equiv \bar{C}$ and ${C}^{(-)}\equiv C$.
The transformation of the gauge field $A^{(+)}_{\mu}$ becomes the usual
gauge transformation with parameter $\theta\,C^{(+)}\equiv
\Lambda^{(+)}$.


\begin{thebibliography}{99}

\bibitem{Hasan}
M. Z. Hasan, C. L. Kane,
Rev.\ Mod.\ Phys.\ \textbf{82}, 3045 (2010).

\bibitem{Qi_Zhang}
X. L. Qi and S. C. Zhang,
Rev.\ Mod.\ Phys.\ \textbf{83}, 1057 (2011).

\bibitem{Ando}
Y. Ando,
J. Phys.\ Soc.\ Jpn.\ \textbf{82}, 102001 (2013).

\bibitem{Kane1}
C. L. Kane, E. J. Mele,
Phys.\ Rev.\ Lett.\ \textbf{95}, 146802 (2005).

\bibitem{Kane2}
C. L. Kane, E. J. Mele,
Phys.\ Rev.\ Lett.\ \textbf{95}, 226801 (2005).

\bibitem{Vozmediano}
A. Cortijo, A. G. Grushin, and M. A. H. Vozmediano,
Phys.\ Rev.\ B \textbf{82}, 195438 (2010).

\bibitem{Bernevig}
B. A. Bernevig and S.-C. Zhang,
Phys.\ Rev.\ Lett.\ \textbf{96}, 106802 (2006).

\bibitem{Frohlich1}
J. Frohlich and T. Kerler,
Nucl.\ Phys.\ B\textbf{354}, 369 (1991).

\bibitem{Frohlich2}
J. Frohlich and A. Zee,
Nucl.\ Phys.\ B\textbf{364}, 517 (1991).

\bibitem{Wen}
X. G. Wen,
Adv.\ Phys.\ \textbf{44}, 405 (1995).

\bibitem{Faddeev-Popov}
L. D. Faddeev and V. N. Popov,
Phys.\ Lett.\ B\textbf{25}, 29 (1967).

\bibitem{BRS}
C. Becchi, A Rouet, and R Stora,
Comm.\ Math.\ Phys.\ \textbf{42}, 127 (1975).

\bibitem{T}
I. V Tyutin, Lebedev Institute preprint N39 (1975).
This work was made widely available as arXvi:0812.0580.

\bibitem{Weinberg}
For a pedagogical introduction to BRST symmetries, see S. Weinberg,
\textit{The Quantum Theory of Fields, Vol. 2, Modern Applications},
Cambridge University Press, New York, 1996.

\bibitem{Polchinski}
For another pedagogical introduction to BRST symmetries, see
J. Polchinski, \textit{String Theory, Vol.1},
Cambridge University Press, New York, 2005.

\bibitem{footnote: Kane-Mele-Hubbard} We emphasize that many-body interactions were neglected in
this paper. We refer the reader to the numerical study of
the Kane-Mele model augmented by a Hubbard interaction by
M. Laubach, J. Reuther, R. Thomale, and S. Rachel,
arXiv:1312.2934.

\end{thebibliography}
\end{document}